\documentclass{nature}

\usepackage{amsmath, amssymb, soul}
\usepackage{graphicx}
\usepackage{pdfpages}
\usepackage{braket}
\usepackage[font={small}]{caption}

\title{Andreev Molecules in Semiconductor Nanowire Double Quantum Dots}

\author{Zhaoen Su$^{1}$, Alexandre B. Tacla$^2$, Mo{\"i}ra Hocevar$^{3,4}$, Diana Car$^5$, S\'ebastien R. Plissard$^6$, Erik P.A.M. Bakkers$^{5,7}$, Andrew J. Daley$^2$, David Pekker$^{1}$, \& Sergey M. Frolov$^1$*}

\begin{document}

\maketitle

\begin{affiliations}
 \item Department of Physics and Astronomy, University of Pittsburgh, Pittsburgh, PA 15260, USA
 \item Department of Physics and SUPA, University of Strathclyde, Glasgow G4 0NG, UK
 \item Universite Grenoble Alpes, F-38000 Grenoble, France
 \item CNRS, Institut Neel, F-38000 Grenoble, France
 \item Department of Applied Physics, Eindhoven University of Technology, 5600 MB Eindhoven, The Netherlands
 \item LAAS CNRS, Universit{\'e} de Toulouse, 31031 Toulouse, France
 \item QuTech and Kavli Institute of Nanoscience, Delft University of Technology, 2628 CJ Delft, The Netherlands
\end{affiliations}
 
\begin{abstract}
Quantum simulation is a way to study unexplored Hamiltonians by mapping them onto the assemblies of well-understood quantum systems\cite{GeorgescuPMP14} such as ultracold atoms in optical lattices\cite{simulation_cold_atom}, trapped ions\cite{simulation_ion} or superconducting circuits\cite{simulation_circuit}. Semiconductor nanostructures which form the backbone of classical computing hold largely untapped potential for quantum simulation\cite{semiconductor_hubbard, singhascience11,BarthelemyAnnPhys13}. In particular, chains of quantum dots in semiconductor nanowires can be used to emulate one-dimensional Hamiltonians such as the toy model of a topological p-wave superconductor\cite{KitaevPU01,SauNatComm12, FulgaNJP13, ZhangNJP16}. Here we realize a building block of this model, a double quantum dot with superconducting contacts, in an indium antimonide nanowire\cite{PlissardNanoLett12}. In each dot, tunnel-coupling to a superconductor induces Andreev bound states\cite{eichlerPRL2007, S_QD_S_prb, ABS_Tarucha_PRL,first_ABS, ABS_Nadya, changPRL2013, spin_resolved_ABS}. We demonstrate that these states hybridize to form the double-dot Andreev molecular states. We establish the parity and the spin structure of Andreev molecular levels by monitoring their evolution in electrostatic potential and magnetic field. Understanding Andreev molecules is a key step towards building longer chains which are predicted to generate Majorana bound states at the end sites\cite{MourikScience2012,Nadj-PergeScience2014}. Two superconducting quantum dots are already sufficient to test the fusion rules of Majorana bound states, a milestone towards fault-tolerant topological quantum computing\cite{AasenPRX16,Shermanarxiv16,Karzigarxiv16}.
\end{abstract}

\begin{figure*}[h!]
\centering
  \includegraphics[width=0.8\textwidth]{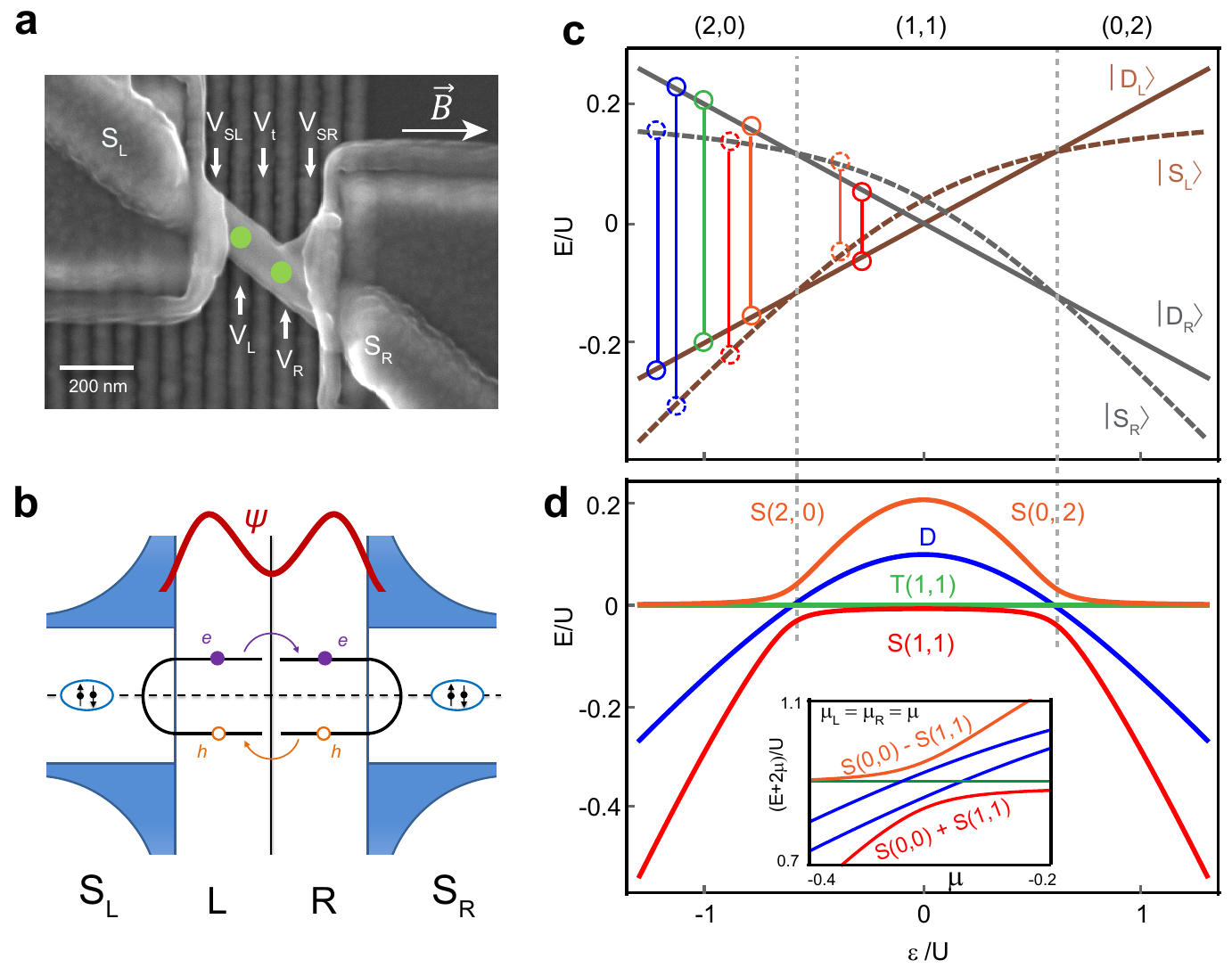}
  \caption{
  \textbf{Superconducting double dot and its energy levels.} 
  \textbf{a,} Scanning electron micrograph of the InSb nanowire device, green circles indicate positions of the two quantum dots.  The direction of magnetic field \textbf{\textit{B}} is indicated by arrow. \textbf{b,} Andreev bound states in two quantum dots coupled by tunneling via the interdot barrier. The red line depicts Andreev molecular wavefunction. \textbf{c,} Spectrum of Andreev states in two quantum dots separated by a large barrier as a function of detuning $\bf{\epsilon}$. On the left(right) dot, the ground states are $\ket {S_{L(R)}}$, $\ket {D_{L(R)}}$ and $\ket {S_{L(R)}}$ with dot occupations 2, 1, 0 (0, 1, 2) respectively. Vertical lines connect levels that hybridize to form molecular states plotted in \textbf{d}. \textbf{d,} Molecular Andreev spectrum of two quantum dots separated by a small barrier as a function of detuning (main panel) and energy level shift (inset). Charge configurations in \textbf{c} and \textbf{d} are labeled in \textbf{c} and separated by dashed lines.
 }
 \label{fig1}
\end{figure*}

In order to realize Andreev molecules we fabricate a device depicted in Fig.\ref{fig1}a. Superconductivity in the InSb nanowire is induced by two NbTiN contacts placed on top of the nanowire\cite{ZhangArxiv2016}, the segments of the wire below the contacts labeled $S_L$ and $S_R$ act as superconducting reservoirs for the left and right dots. The reservoirs are characterized by the induced gap $\Delta \sim 400~\mu$eV. We use voltages on five electrostatic gate electrodes placed under the nanowire to define the two quantum dots. Voltages on the two outer gates set the couplings $\Gamma_L$ and $\Gamma_R$ to the superconducting reservoirs. Gate voltages $V_L$ and $V_R$ control the chemical potentials on the left and right dots. The middle gate labeled $V_t$ controls the coupling $t$ between the dots. While all couplings are tunable in a wide range, here we focus on the regime where the system is approximately left/right symmetric, and with $\Gamma_L, \Gamma_R$ $>$ t. In this regime the two dots are strongly coupled to their respective superconducting reservoirs and weakly coupled to each other. The charging energy on each dot $U\sim 1-2$ meV $>$ $\Delta$ thus the dots can be filled by electrons one at a time rather than in Cooper pairs.

In superconductor-semiconductor hybrid structures, electrons arriving from a semiconductor with energies below the superconducting gap are prohibited from entering the superconductor and are reflected back into the semiconductor as quasiholes via Andreev reflection\cite{BlonderPhysRevB1982}. Through this mechanism, an electron-hole standing wave, known as an Andreev bound state, can form in the semiconductor (Fig.~\ref{fig1}b). In a single quantum dot, Andreev bound state spectrum consists of a spin-singlet state (S) which is a superposition of 0 and 2 electrons on the quantum dot, and two doublet states $D_\uparrow$ and $D_\downarrow$, both of which correspond to a single electron on a quantum dot either in the spin up or spin down state. In Fig.~\ref{fig1}c, we depict the Andreev spectra of two decoupled quantum dots along the energy level detuning axis, meaning that the electrostatic energies on the two dots are changed in the opposite directions. From negative to positive detuning, the left dot is occupied with 2, 1 and 0 electrons, while the right dot is occupied with 0, 1, and 2. In the (2,0) and (0,2) double dot  configurations, singlet states on both dots are lower in energy than doublets. In the (1,1) configuration, doublets are the ground states.

When the two dots are tunnel-coupled, each of the states on one dot will hybridize with each of the states on the other dot (Fig. ~\ref{fig1}b).
The resulting Andreev molecular spectrum is depicted in Fig.\ref{fig1}d. The new singlet states are S(0,2), S(2,0) and S(1,1): these three states hybridize at their degeneracy points due to tunnel coupling.  The four doublet states hybridized of D(0,1) and D(1,0), D(2,1) and D(1,2) are nearly degenerate at zero field and are designated as D in Fig.\ref{fig1}d and are always the excited states. When the chemical potentials $\mu_L$ and $\mu_R$ on the left and right dots are tuned along the energy shift axis, such that $\mu_L=\mu_R=\mu$ the double dot can transition from (0,0) to (1,1) configuration. In this case, S(0,0) and S(1,1) are hybridized by superconducting correlations (Fig. 1d(inset)).

\begin{figure*}[h!]
\centering
  \includegraphics[width=0.9\textwidth]{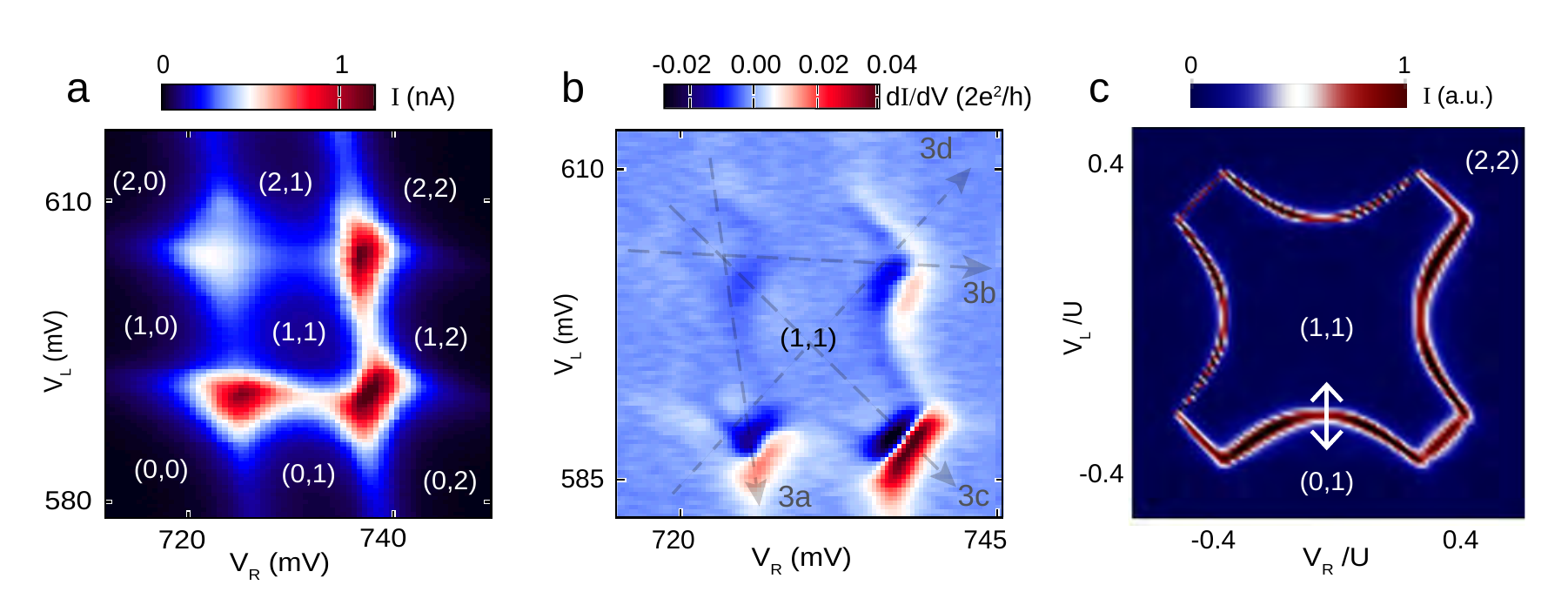}
  \caption{
  \textbf{Charge stability diagram.}
\textbf{a,} Current at  $V_{bias}$ = 200 $\mu$V. Parities of double dot configurations are indicated in brackets. \textbf{b,} Differential conductance $dI/dV$ over the same gate voltage range as in {\bf a}. Dashed lines indicate the cuts that correspond to panels in Fig.~\ref{fig3}. \textbf{c,} Numerically computed current at low interdot tunneling. The lower arc-shaped transport peak due to resonance between Andreev states in (1,1) and (0,1) is marked with an arrow. (See supplementary information for details). 
 }
 \label{fig2}
\end{figure*}

A new type of levels appears below the gap 
in a double quantum dot: the three triplet states $T_+$(1,1)=$(\uparrow,\uparrow)$, $T_-$(1,1)=$(\downarrow,\downarrow)$ and $T_0$(1,1)=$(\uparrow,\downarrow)+(\downarrow,\uparrow)$ trace back to the symmetric combinations of single-dot doublet states. T(0,2) and T(2,0) are above the induced gap due to the large orbital energy and thus they do not correspond to bound Andreev states. In experiment, source-drain voltage bias $V_{bias}$ is applied between $S_L$ and $S_R$, thus the chemical potentials in the two superconductors are experimentally tunable. This influences the Andreev spectra because Andreev levels are pinned to the energy interval of width 2$\Delta$  around the chemical potential (see supplementary information).

Measurements below are focused on a double dot stability diagram presented in Fig.\ref{fig2}a (see supplementary information for expanded diagram). Four degeneracy points are observed at which the current has a local maximum. The upper-left maximum of current is lower than the other three. In reverse $V_\text{bias}$, the lower-right maximum has the lowest current. This is due to spin blockade which occurs between (1,1) and (0,2) or (2,0) double dot  states due to Pauli exclusion (see supplementary information for further evidence)\cite{insb_spin_blockade}. Spin blockade is a manifestation of hybridized quantum states on the two dots, and it allows us to identify and label the parity of nine configurations in Fig.\ref{fig2}a. The dot occupations are higher than their parities. 

\begin{figure*}[h!]
\centering
  \includegraphics[width=0.9\textwidth]{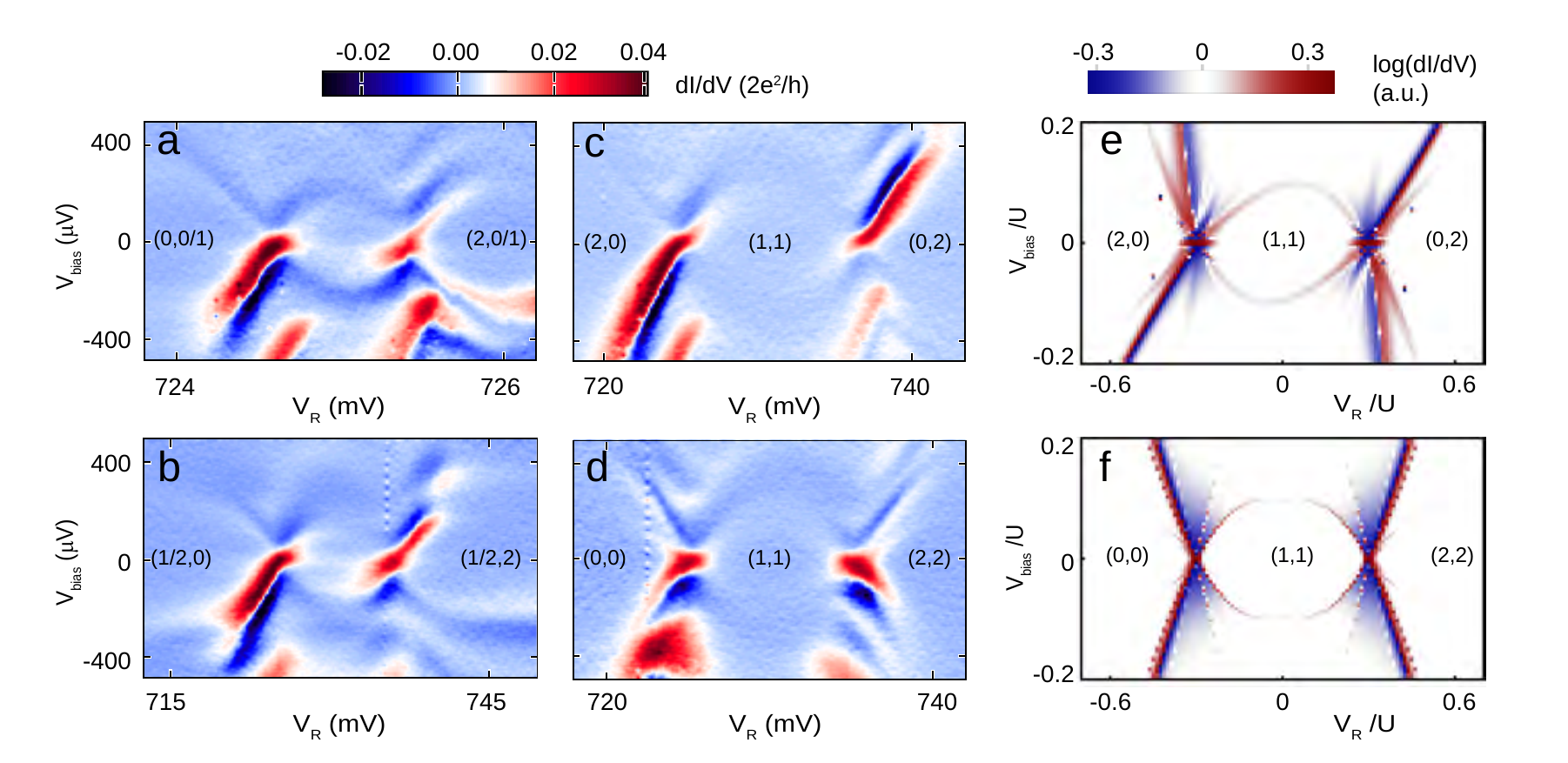}
  \caption{\textbf{Bias spectroscopy of Andreev molecular states.}
 \textbf{a-d,} Source-drain bias spectroscopy along various cuts depicted by the dashed-line arrows in Fig.~\ref{fig2}b, i.e., right (left) dot is fixed to 0/1 (1/2) degeneracy point and left (right) dot is swept in \textbf{a} (\textbf{b}), along detuning in \textbf{c} and along energy level shift axis in \textbf{d}. Both $V_L$ and $V_R$ are tuned, and $V_R$ is used to denote the x-axis. Charge configurations indicated in brackets in each region. \textbf{e} and \textbf{f}, Numerically computed differential conductance along detuning (\textbf{e}) and energy level shift axis (\textbf{f}). 
 } 
 \label{fig3}
\end{figure*}

In differential conductance the double dot stability diagram is defined by arc-shaped resonances that connect the degeneracy points (Fig.\ref{fig2}b).  Numerical simulation of transport confirms that these arcs are due to resonances between Andreev states with different double dot configurations. For instance, the resonance between Andreev states in (1,1) and (0,1) yields the lower arc-shaped transport peak depicted in Fig.\ref{fig2}c.
The same mechanism produces loop-like resonances in gate vs. bias scans (Fig.\ref{fig3}). These resonances appear when either dot is fixed at a  degeneracy point and the other dot is swept (Fig.\ref{fig3}a-b). The similarity between Fig.\ref{fig3}a and Fig.\ref{fig3}b indicates that the system is symmetric, and Andreev reflection occurs both from the left and the right leads. Loop-like resonances are also observed when the energy levels on the two dots are tuned simultaneously (Fig.\ref{fig3}c-d). These resonances are contained within $\pm ~400  ~\mu V$ and are accompanied by copies in negative differential conductance. 

The observed Andreev loops are closed, i.e. the conductance resonances reach zero bias. This is counter-intuitive given that both leads of the system are superconductors and thus an energy gap is expected around zero bias\cite{first_ABS,S_QD_S_prb}. We ascribe the observed low bias transport to sub-gap quasi-particles that enable single-particle transport. When this effect is included in the numerical model, simulations reproduce the closed loops and negative differential conductance (Figs. \ref{fig3}e-f). We model each lead as being composed of two parts: a conventional superconductor with a hard superconducting gap and a normal Fermi gas with gapless excitations. The electrochemical potentials of the normal and the superconducting parts are pinned together at the value set by the voltage applied to the physical lead. In our model, Andreev reflection off the superconducting part results in the formation of Andreev molecules. The normal part induces transitions between the Andreev molecular states (see supplementary information for details).

\begin{figure*}[h!]
\centering
  \includegraphics[width=0.9\textwidth]{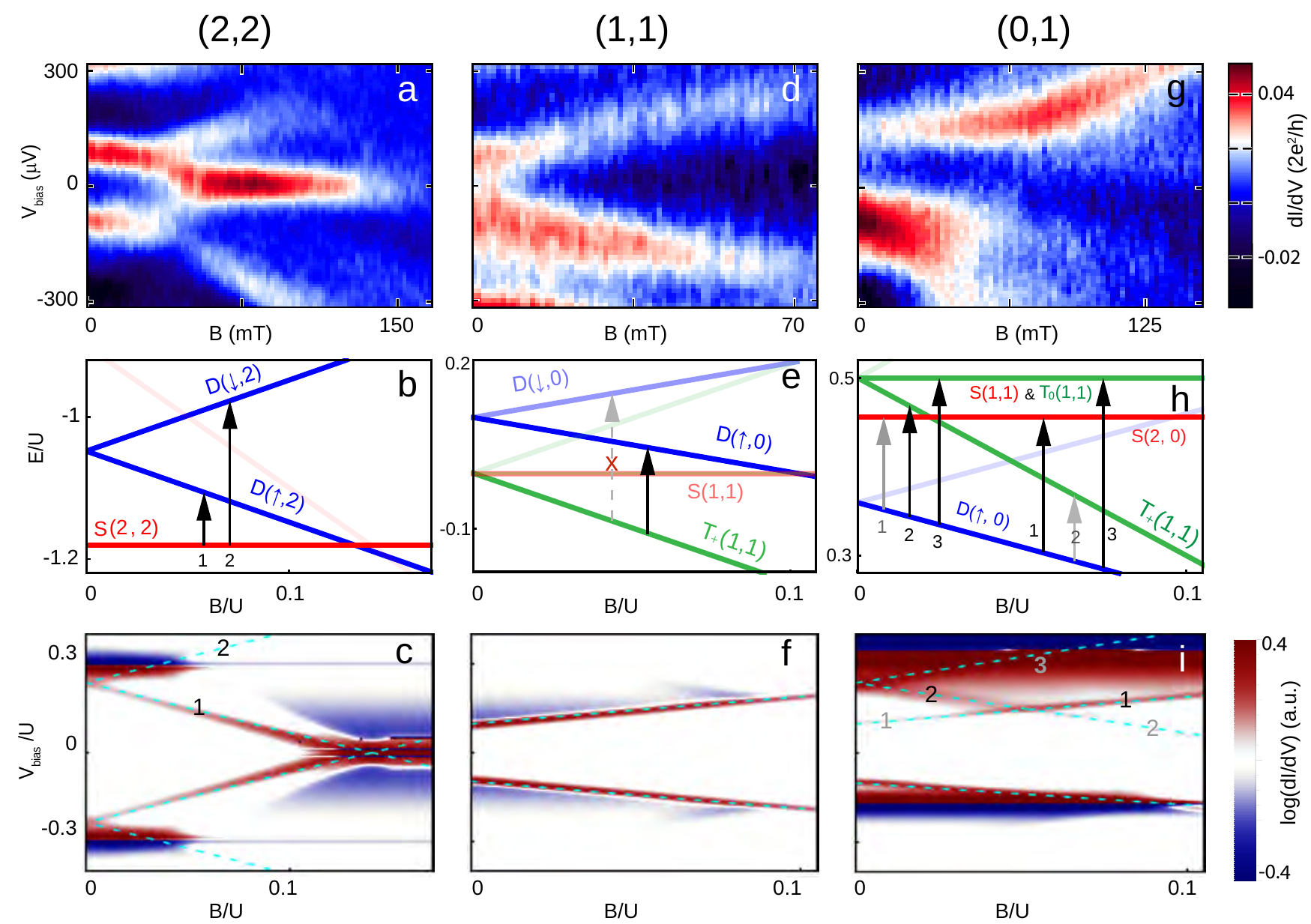}
  \caption{
  \textbf{Magnetic field evolution of Andreev molecular states.}
\textbf{a, d, g}, Bias spectroscopy of Andreev resonances as a function of magnetic field for $(2,2)$, $(1, 1)$ and $(1, 0)$ double dot configurations.  \textbf{b, e, h}, Numerically computed spectra of Andreev molecular states as a function magnetic field for $V_\text{bias}=0$. The black (gray) arrows and numbers label the allowed transitions in the simulated spectra (\textbf{b, e, h}) and the associated high (low) conductance resonances in the numerical $dI/dV$ transport plots (\textbf{c, f, i}). In \textbf{e} the crossed dashed arrow labels the forbidden transition between $T_+(1,1)$ and $D(\downarrow,0)$. The light blue dashed lines in \textbf{c, f, i} plot the bias voltage at which the levels on the dots come into resonance. The $dI/dV$ plots use the same model parameters as in the spectrum plots. 
 } 
 \label{fig4}
\end{figure*}

We investigate the spin structure of Andreev molecular states by monitoring the evolution of subgap transport features in magnetic field. In Fig.~\ref{fig4}a we plot differential conductance as a function of magnetic field and source drain bias for a double quantum dot in the (2,2) configuration. At zero magnetic field we observe two peaks, one at positive bias and one at negative bias. The application of magnetic field results in the splitting of both peaks. Two of the peaks move to higher bias towards the gap edge, while the other pair meets at zero bias. The two merged resonances stick to zero bias at finite field. This effect has been investigated as a signature of Majorana fermions\cite{MourikScience2012}. Here, given the narrow range of field over which the zero-bias peak is observed, we associate it with level repulsion from the gap edge or from other subgap states\cite{spin_resolved_ABS}. By comparing measurements to numerical spectra and transport calculations, we assign the peaks to the transitions between the $S(2,2)$ ground state and the $D(\uparrow,2)$ and the $D(\downarrow,2)$ excited states (Fig.\ref{fig4}b,c). Magneto-transport of the double quantum dot system in the (0,0), (0,2) and (2,0) configurations is qualitatively the same as in the (2,2) configuration (see supplementary information).

In the (1,1) configuration only a single pair of differential conductance peaks is observed at all fields, one at positive and one at negative bias (Fig.~\ref{fig4}d). Both peaks shift to higher bias at higher magnetic fields. The explanation for this behavior originates in the Andreev molecular level structure depicted in Fig.~\ref{fig4}e. The low energy manifold consists of $S(1,1)$ ground state that is almost degenerate with the three triplet states $T_+$, $T_0$, $T_-$. At finite field $T_+$ plunges below the $S(1,1)$ and becomes the ground state. Transitions from this triplet state are allowed only to the doublet states $D(\uparrow,0)$, while transitions to $D(\downarrow,0)$ are strongly suppressed because they involve an additional spin flip. Both states $T_+$ and $D(\uparrow,0)$ shift to lower energies with magnetic field, but the triplet states shifts with $g\mu_B B$ while the doublet states shifts with $g\mu_B B/2$, thus the energy difference between them grows with field. Transport calculations using our detailed model confirm this picture (Fig.~\ref{fig4}f).

Odd total parity configurations (0,1), (1,0), (2,1) and (1,2) offer a richer variety of transport behavior (Fig.~\ref{fig4}g, and supplementary information). The common features include asymmetry with respect to bias and kinks in the conductance peaks at which the effective g-factor increases. In some regimes we also observe the magnetic-field induced splitting of conductance peaks into as many as three sub-peaks.
In Fig.~\ref{fig4}h we plot the Andreev molecular spectrum in the (0,1) configuration as a function of magnetic field at zero bias. While  $D(0,\uparrow)$ is the well-separated ground state at finite field, there are two singlet states ($S(0,2)$ and $S(1,1)$) and two triplet states ($T_+$ and $T_0$) that can contribute to transport (transport via the state $T_-$ requires a spin flip and is therefore suppressed). Numerically computed transport demonstrating both a kink feature as well as the asymmetry with respect to bias, is plotted in Fig.~\ref{fig4}i. The model indicates that the origin of the kink feature is that as $B$ increases the $D(\uparrow,0) \rightarrow T_+(1,1)$ transition (labeled 2 in Fig.~\ref{fig4}h,i) becomes dimmer while the $D(\uparrow,0) \rightarrow S(2,0)$ transition (labeled 1 in Fig.~\ref{fig4}h,i) becomes brighter. The dimming and brightening of the transitions is associated with proximity to the interdot resonances that occurs at higher bias. The model shows that the origin of the bias asymmetry is a combination of two factors. First, the energies of the Andreev molecular states are bias dependent, as electrons participating in these states are delocalized between the quantum dots and the leads. Second, the gate setting (V$_L$, V$_R$) is asymmetric, meaning such gate setting results in different parities on the two dots.

\begin{methods}
The nanowires (diameter 100 nm) are grown in the 111 crystal orientation by metalorganic vapor phase epitaxy from gold catalysts, as described in ref.12
. Local gate electrodes (pitch 60 nm) are defined by electron beam lithography and electron beam evaporation of Ti(5 nm)/Au (10 nm) on thermal silicon oxide. The gate electrodes are then covered by Atomic layer dposition grown HfO$_2$ (10 nm). Single InSb nanowires are transferred by a micromanipulator. The superconducting contacts are Ti/NbTi/NbTiN (5/5/150 nm). Prior to sputtering the nanowires are passivated in ammonium sulfide to remove the native oxide. 

The measurements are performed at 35 mK in a dilution refrigerator. A d.c. voltage bias is applied to the left superconducting lead (S$_L$) and the current from the right superconducting lead (S$_R$) to the ground is measured by a current amplifier.  To measure the differential conductance, a standard lock-in technique is used (77 Hz, 5 $\mu$V).
\end{methods}
\newpage

\newpage

\newpage

\bibliographystyle{unsrt}
\bibliography{Ref.bib}

\begin{addendum}
 \item We thank R. Aguado, A. Akhmerov, S. De Franceschi, E. Lee, V. Liu for valuable discussions. Work is supported by Charles E. Kaufman Foundation (S.M.F. and D. P.), NSF DMR-125296, ONR N00014-16-1-2270 (S. M. F.) and AFOSR FA9550-12-1-0057 (A. D.).
 
 \item[Author Contributions] D.C., S.P. and E.B. grew InSb nanowires. Z.S. and M.H. fabricated devices. Z.S. and S.F. performed the measurements and analyzed data. A.T., A.D. and D.P. performed numerical simulations. All authors wrote the manuscript.

 \item[Correspondence] Correspondence and requests for materials should be addressed to S.M.F.~(email: frolovsm@pitt.edu).
\end{addendum}



\section*{Supplementary material}

## ORGANIZATION OF THE SUPPLEMENT

The supplement consists of three sections. In section I we provide a set of figures with additional experimental data and results of theoretical calculations. In section II we describe the minimal model of Andreev molecules at zero source-drain bias. Finally, in section III we provide a detailed description of our transport model.

## I. SUPPLEMENTARY EXPERIMENTAL DATA AND THEORETICAL RESULTS

List of Supplementary Experimental Figures:

Fig. S1 : double dot parity stability diagram in the expanded gate range.

Fig. S2 : study of spin blockade in the double dot configuration studied in the main text.

Fig. S3 : magnetic field evolution of the stability diagram.

Fig. S4 : additional gate vs. gate and bias vs. gate subgap resonance maps.

Fig. S5 : magnetic field dependence of the full 3x3 configurations around (1,1).

Fig. S6 : magnetic field dependence of subgap resonances near (2,1)↔(1,2) degeneracy point.

Fig. S7: Bias spectroscopy at zero field and at finite field.

Fig. S8 : conductance maps in gate and bias for strongly coupled dots.

Fig. S9 : magnetic field evolution of subgap resonances in strongly coupled dots.

Fig. S10 : high bias spectroscopy on a single dot.

List of Supplementary Theoretical Figures:

Fig. S11 : simulated spectra of the S-QD-QD-S system.

Fig. S12 : simulated parity diagram and bias vs. gate subgap resonance.

Fig. S13 : simulated magnetic field evolution of the full 3x3 configurations around (1,1).

## A. Expanded double dot stability diagram

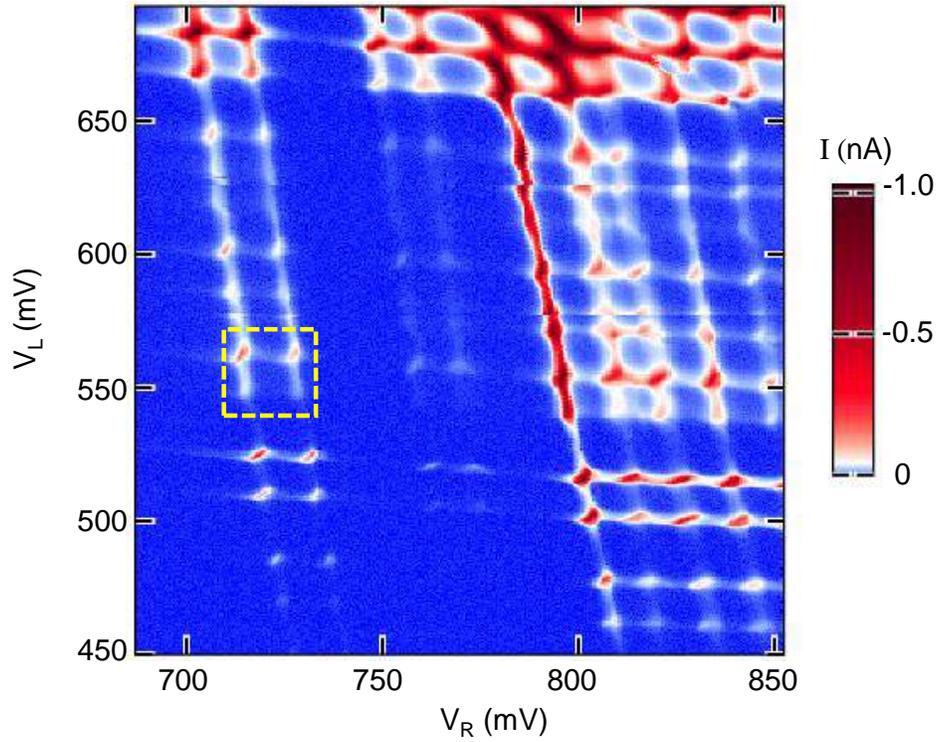

FIG. S1. Double dot stability diagram at the bias of -200 $\mu$V and at zero magnetic field in a large gate voltage range. Measurements in the main text focus on the window enclosed by the yellow square. We note that the left dot has a charge instability, meaning that the features shift along the vertical axis spontaneously. The instability, however, is tractable and we are able to tune $V_L$ to compensate the voltage change when it occurs. This results in mismatch in $V_L$ ranges in conductance maps taken at different times.

## B.  Spin Blockade

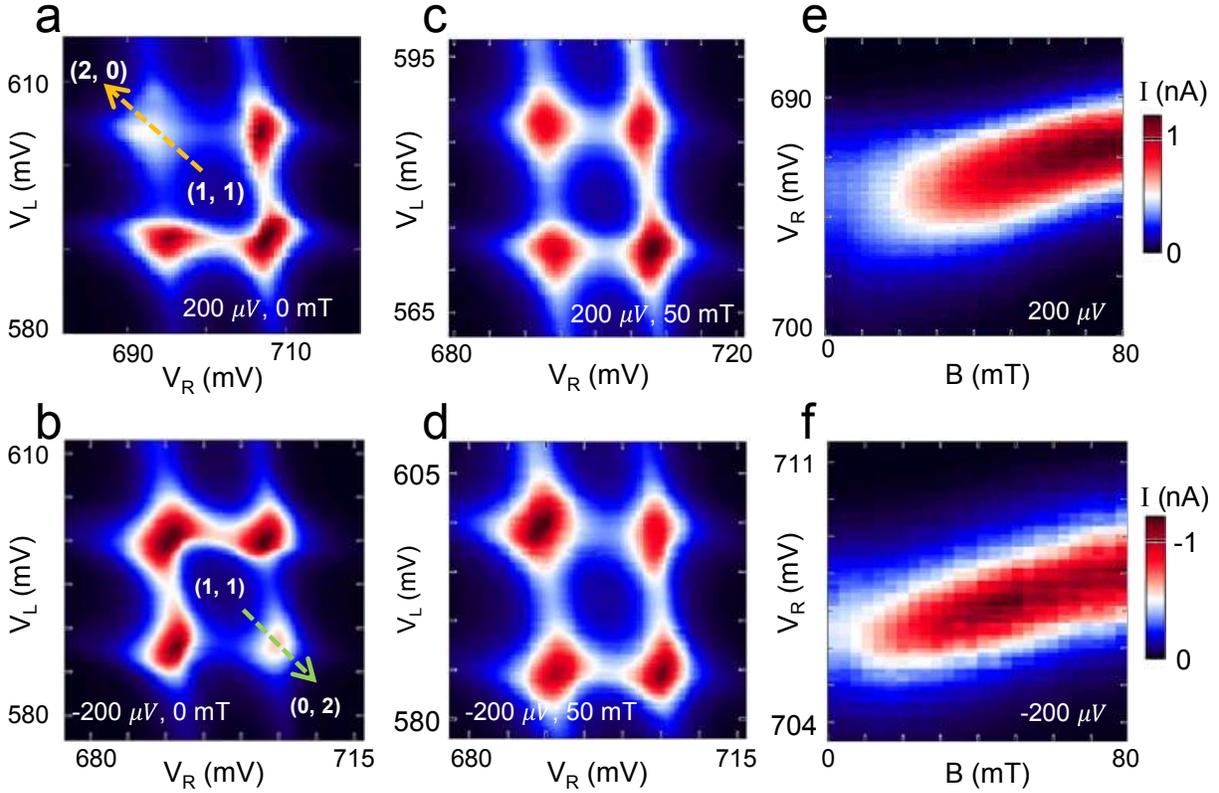

FIG. S2. **a**, At positive bias of 200 $\mu$V the (1, 1)↔(2,0) degeneracy point has smaller current. **b**, At the reverse bias the (1, 1)↔(0,2) degeneracy point has the smaller current. **c, d**, Spin blockade is lifted by a magnetic field of 50 mT, with all four degeneracy points showing similar current levels. In **e, f**, gates are swept through the spin blockade region (arrows in **a** and in **b**) as magnetic field is stepped. The current has a zero-field dip, consistent with spin blockade reported previously for InSb double quantum dots[1].

## C. Magnetic field evolution of the charge stability diagram

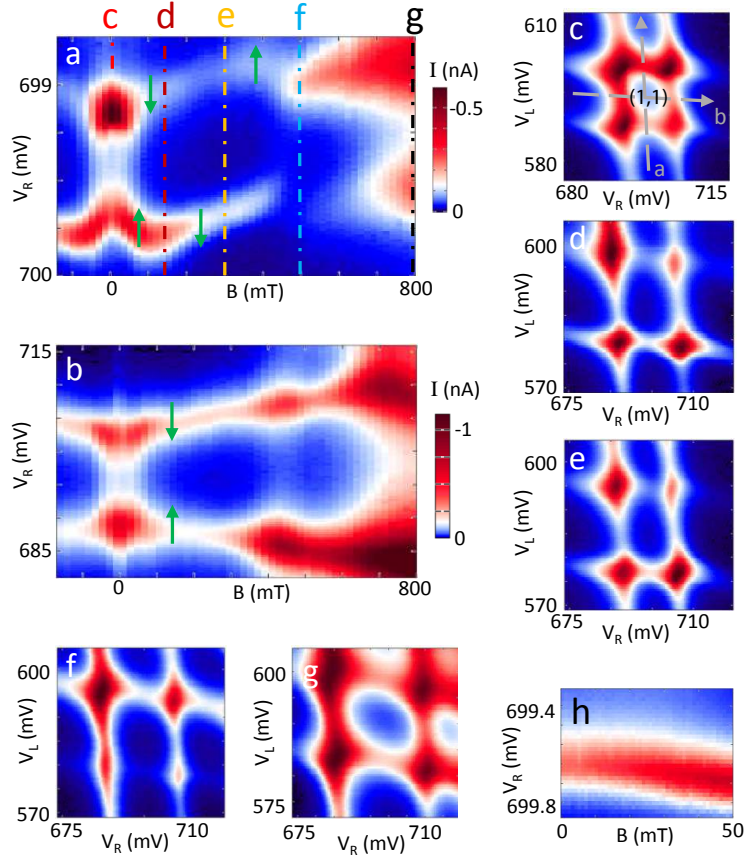

FIG. S3. (**a-b**), The evolution of current along the line cuts a and b shown in panel **c**, $V_{bias} = -200$ $\mu$V. For B $\leq$ 100 mT, the high current lines that define the (1,1) region move apart. At higher fields, the lines exhibit kinks. They are are consistent with previously reported ground state spin transitions in InSb nanowire quantum dots[2]. Spins of the electrons added to the dot are depicted by arrows along the high current lines. **c-g**, The double dot stability diagrams at 0, 140, 300, 500, 800 mT, $V_{bias} = -200$ $\mu$V. We notice that the higher current lines connected the spin blockade points (lower branch of **a** and higher branch of **b**) also undergo lower current at zero field. It is specially demonstrated with a zoomed-in scan of the lower branch from 0 to 50 mT (**h**). The color bar of **b** is on its right side. The other plots share the color bar on the right side of **a**.

## D. Complimentary data on Andreev Molecular States

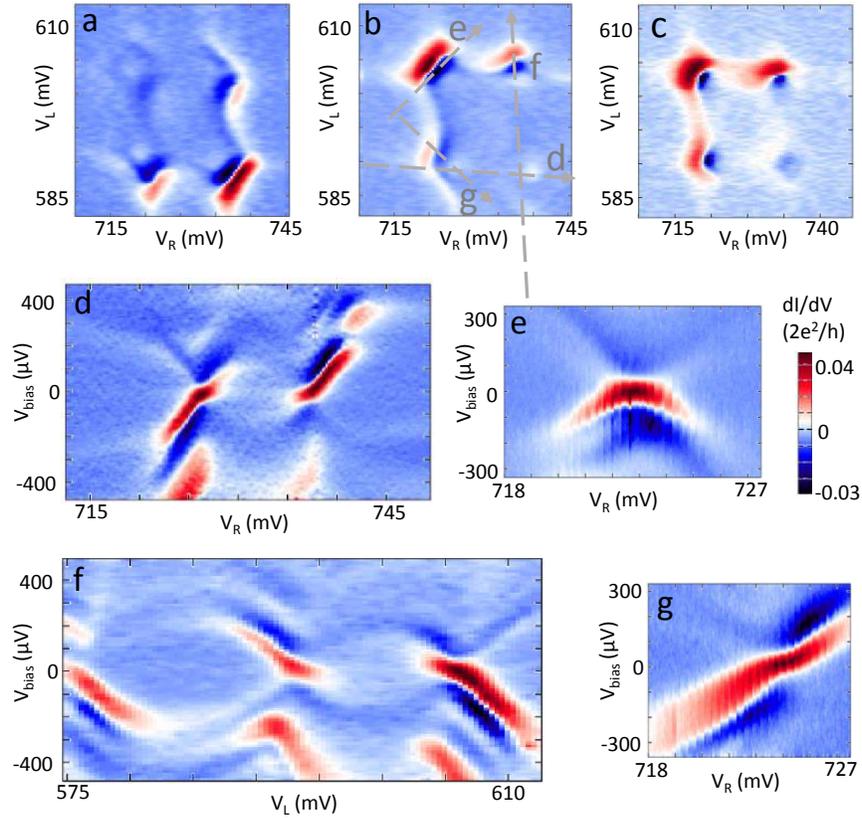

FIG. S4. Spectroscopy measurements of Andreev molecular resonances complimentary to Fig.2 and Fig.3 of the main text. **a-c**, Gate vs. gate differential conductance diagrams at the bias of (a) 200 $\mu$V, (b) -200 $\mu$V and (c) -100 $\mu$V. **d-g**, Source-drain bias spectroscopy scans along various cuts depicted by the arrows in **b**. **d**, With the left dot fixed at its degeneracy point. **f**, With the right dot fixed at its degeneracy point. Here we extend the plot to display adjacent Andreev loops. **e**, Along the symmetric cut through the (1,0)↔(2,1) degeneracy point. **g**, Along the detuning cut through the (1,0)↔(0,1) degeneracy point.

7

## E. Spin map

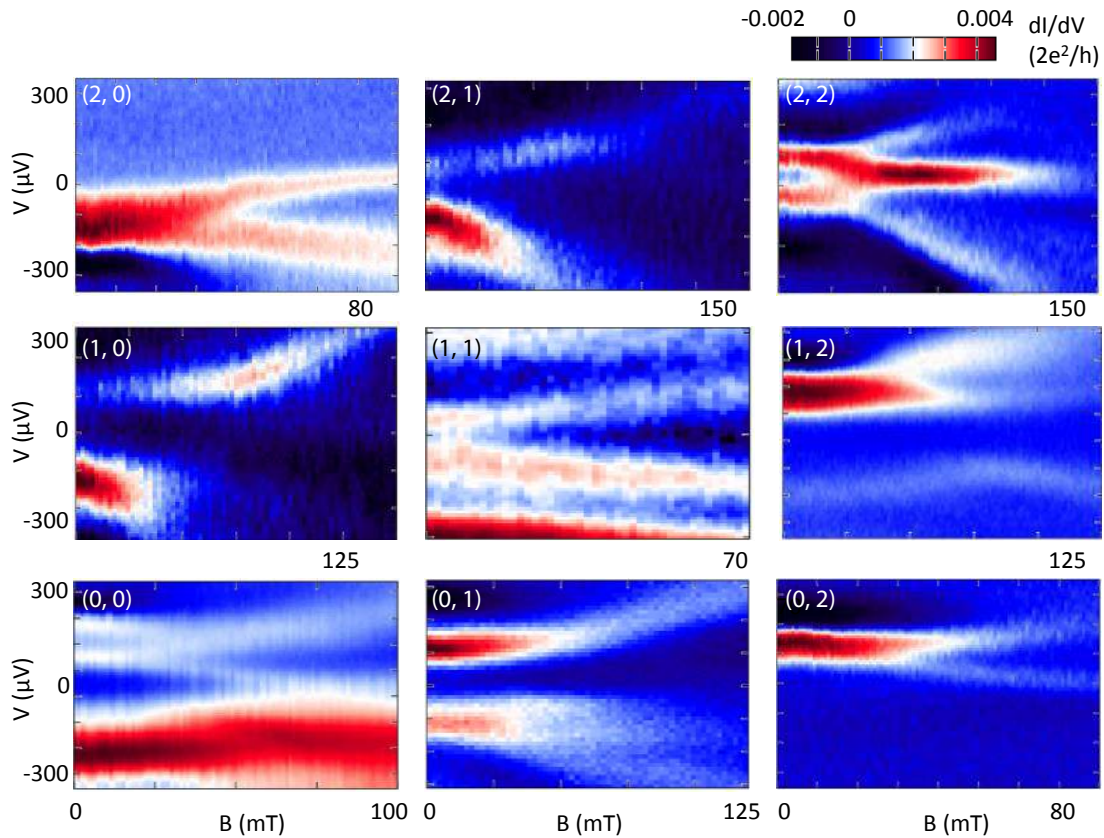

FIG. S5. The magnetic field dependence of the subgap resonances in the 3x3 double dot configurations around (1,1). Within each configuration, the magnetic field dependence of the resonances is measured at a few fixed gate voltages. Scans in each configuration demonstrate consistent magnetic field behavior and one typical scan is presented here. Note that the positive (negative) bias branch in (2,0) ((0,2)) is missing, which is a consequence of spin blockade and asymmetric couplings.

## F. Detailed magnetic field data in the (2,1) configuration

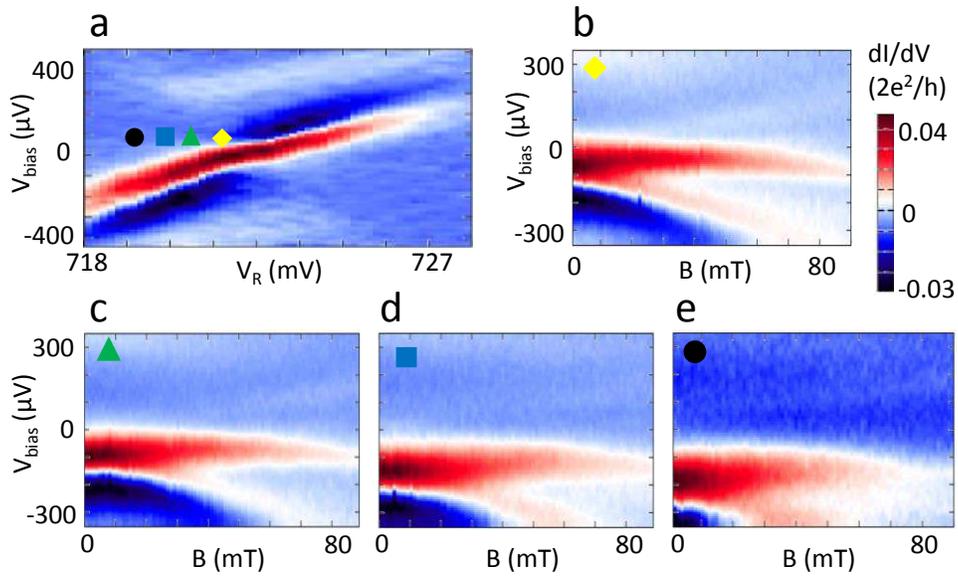

FIG. S6. The magnetic field dependence in (2,1) at various spots away from the degeneracy point. **a**, The spectroscopy along the detuning cut through the (2,1)↔(1,2) degeneracy point. **b-e**, A series of bias vs. field scans in (2,1) at spots farther and farther away from the degeneracy point, depicted by the yellow diamond, green triangle, blue square and black dot marked in panel **a**. This shows that qualitatively the same features are observed within the same quadrant of the double dot stability diagram in Fig.S5.

## G. The source-drain bias spectroscopy at zero and finite fields

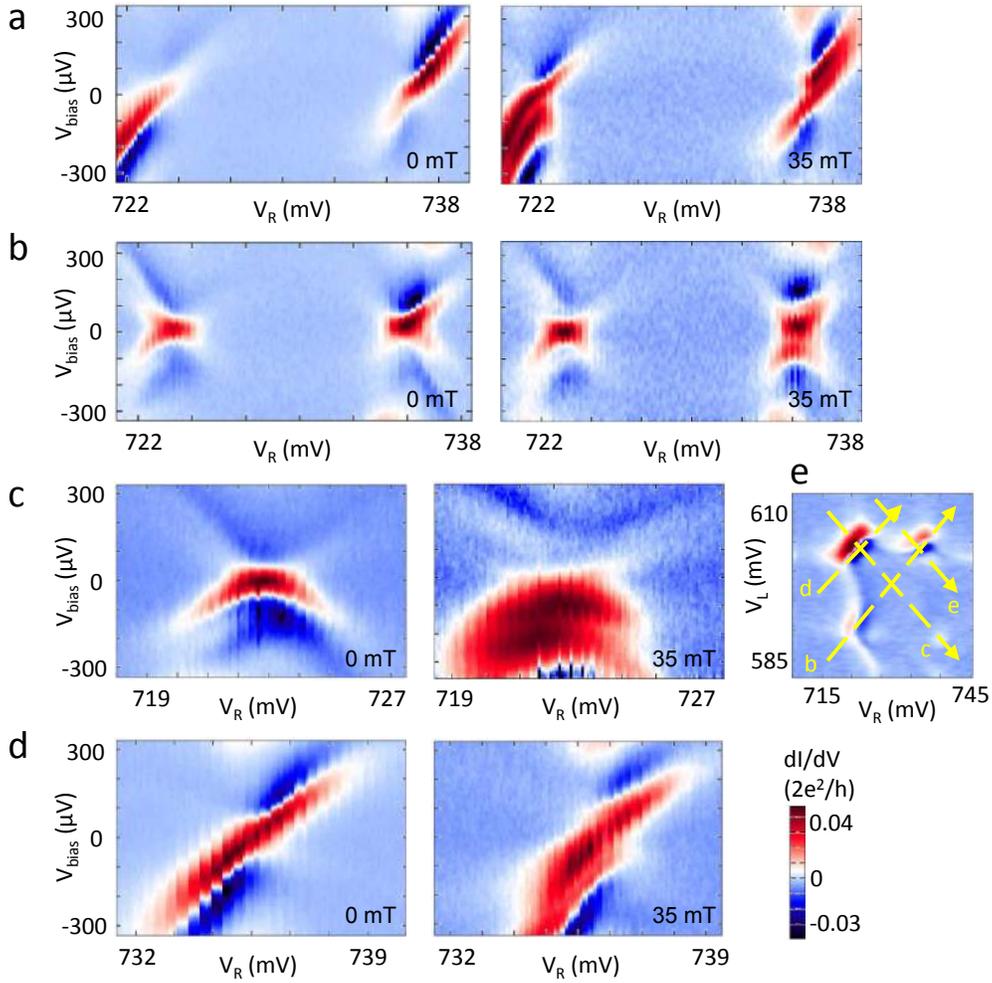

FIG. S7. **a-d**, The spectroscopy measurements at 0 mT (left) and 35 mT (right) along different cuts depicted by the yellow arrows in **e** which is the same as Fig. S4**b**. **a-b**, For both bias directions, resonances are doubled at finite field in the (even, even) configurations; while in (1,1) a resonance does not split. **c**, Along the energy shift cut through the (2,0)↔(1,1) degeneracy point. **d**, Along the detuning cuts through (2,1)↔(1,2) degeneracy point.

## H. Strong interdot coupling regime

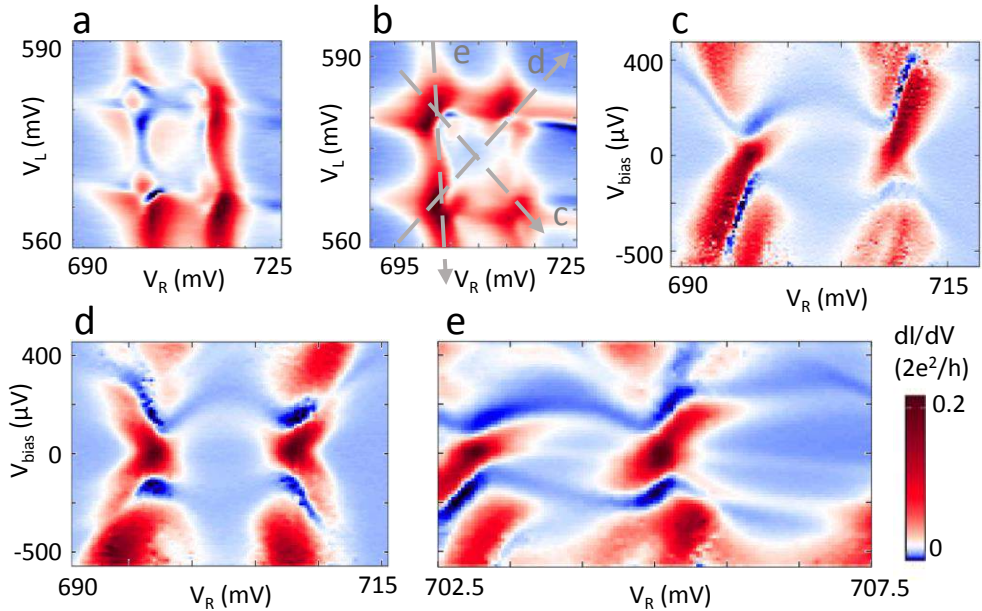

FIG. S8. Data in this plot and Fig.S9 is obtained after the device has been re-tuned to lower the interdot barrier by adjusting gate $V_t$ from $-685$ mV to $-635$ mV, and all other gates to keep the same dot occupations. $V_{\Gamma L}$ and $V_{\Gamma R}$ remain almost the same. **a-b**, The gate vs. gate differential conductance diagrams at biases of 200 $\mu$V and -200 $\mu$V. **c** and **d**, The spectroscopy scans along the detuning and energy shift cuts. **e**, scan the left dot when the right dot is fixed at the degeneracy point. The spectra are qualitatively similar to those in the weaker interdot coupling regime. However, owing to the stronger interdot coupling, a few features which are less clearly resolved in the weaker interdot coupling regime are more pronounced here. First, the resonances away from the degeneracy points are clearly visible. Second, the close-to-zero-bias resonances, that appear within the loop-like resonances are resolved. These resonances appear due to proximity to charge degeneracy in the right dot, they are further discussed in Fig.S11.

## I. The spin map in the stronger interdot coupling regime

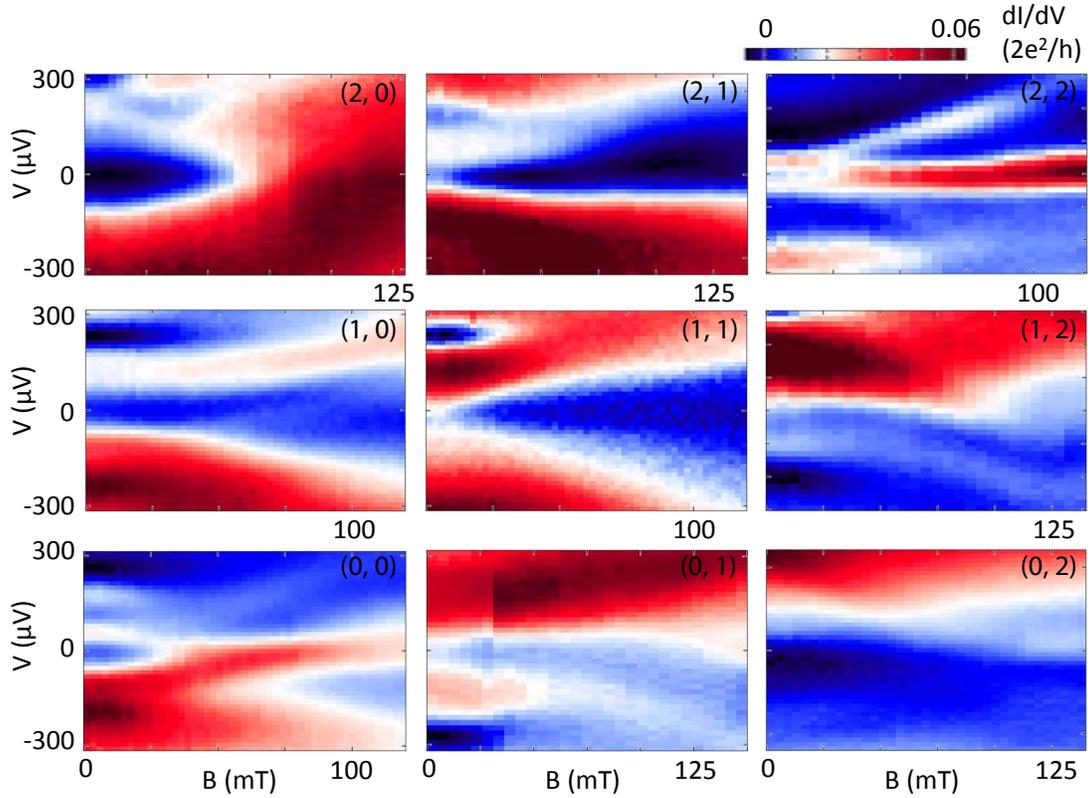

FIG. S9. Spin map in the stronger interdot coupling regime ($V_t = -635$ mV). Magnetrospectroscopy is performed in the 3x3 double dot configurations around (1,1) in the stronger interdot coupling regime. In each configuration a typical example is presented. Although the resonances are more broadened, qualitatively similar magnetic field dependencies are observed here as in Fig. S5. That is, resonances don't split and moves away from zero bias in (1,1), resonances split in the (even, even) configurations, and resonances exhibit kinks at positive (negative) bias and broadened structures at negative (positive) bias in (1,0) and (2,1) ((0,1) and (1,2)).

## J. High bias scans revealing multiple loops of single dots

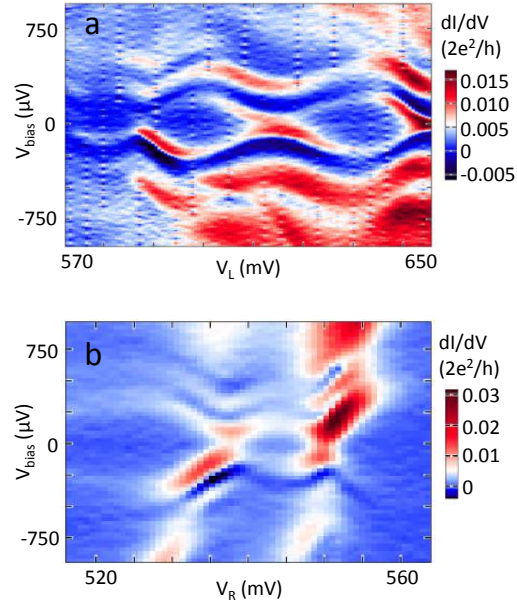

FIG. S10. In **a** we show the spectroscopy of the left single dot when the right dot is removed. The system is therefore a single quantum dot coupled to a superconducting lead and probed by the other superconducting lead that is weakly coupled to the dot. Besides the lowest bias loop-like resonances, we observe copies of these resonances at higher biases. They demonstrate non-linear gate dependence of the same pace of the lowest bias loops. Both lowest bias loops and higher bias resonances are followed by negative differential conductance peaks. These features are observed in the right single dot as well (**b**). We note that the copies are present only in systems with both leads superconducting. When an Andreev bound state is probed by a normal lead, only one loop is observed within the gap. The copies of the Andreev states may originate from quasiparticle peaks in the tunneling probes and multiple Andreev reflections. They are being addressed in a future publication. In this paper we focus on the resonances closest to zero bias.

## K.  Simulated spectra of the S-QD-QD-S system

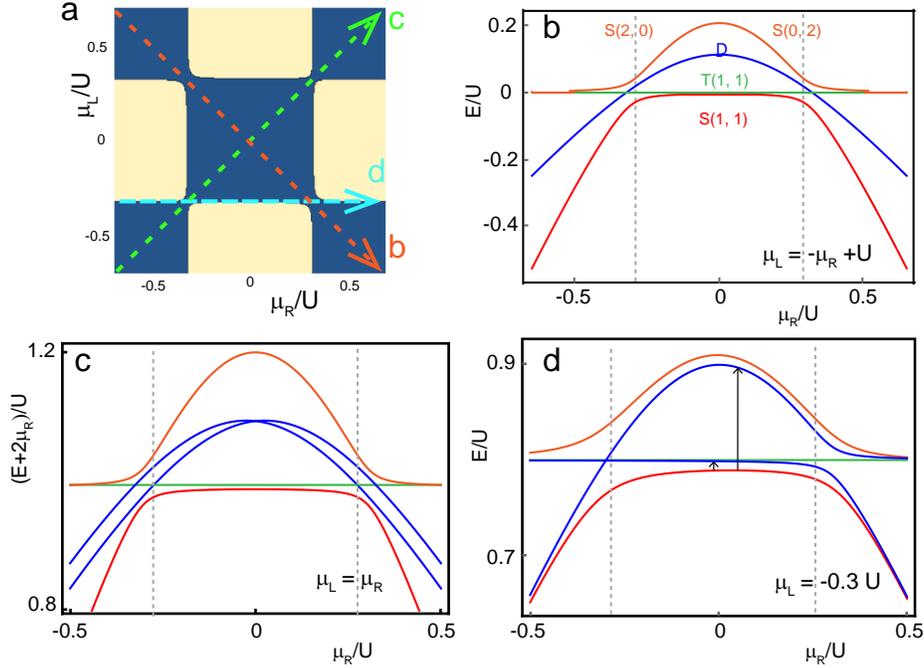

FIG. S11.  Simulated spectra of the system at zero bias. Throughout the paper, the simulated spectra and transport plots use the same parameters, i.e., Andreev reflection matrix element $\tau_{L/R}/U = 0.4$, inter-dot tunneling matrix element $t/U = 0.01$ in terms of the on-site interaction strength $U$. **a**, The simulated diagram showing the ground state parities as a function of left and right dot chemical potential ($\mu_L$ and $\mu_R$). Blue regimes have total parities of even and yellow regimes have total parities of odd. **b-d**, Bias scans along the cuts depicted in **a**. These spectra explain the corresponding bias vs. gate subgap conductance measurements. We notice that, interestingly, in the measurements where one dot is fixed at the degeneracy point (Fig.3**a-b** and Fig.S4**d,f**), additional close-to-zero-bias resonances are observed inside the loop-like structure. They can be explained by **d** where besides the transition from the ground state to the higher doublet state (tall arrow), there exist the transition to the lower doublet state which requires little energy (short arrow). They are associated with adding or removing an electron in the degenerate dot and therefore they cost little energy.

## L. Simulated transport plots

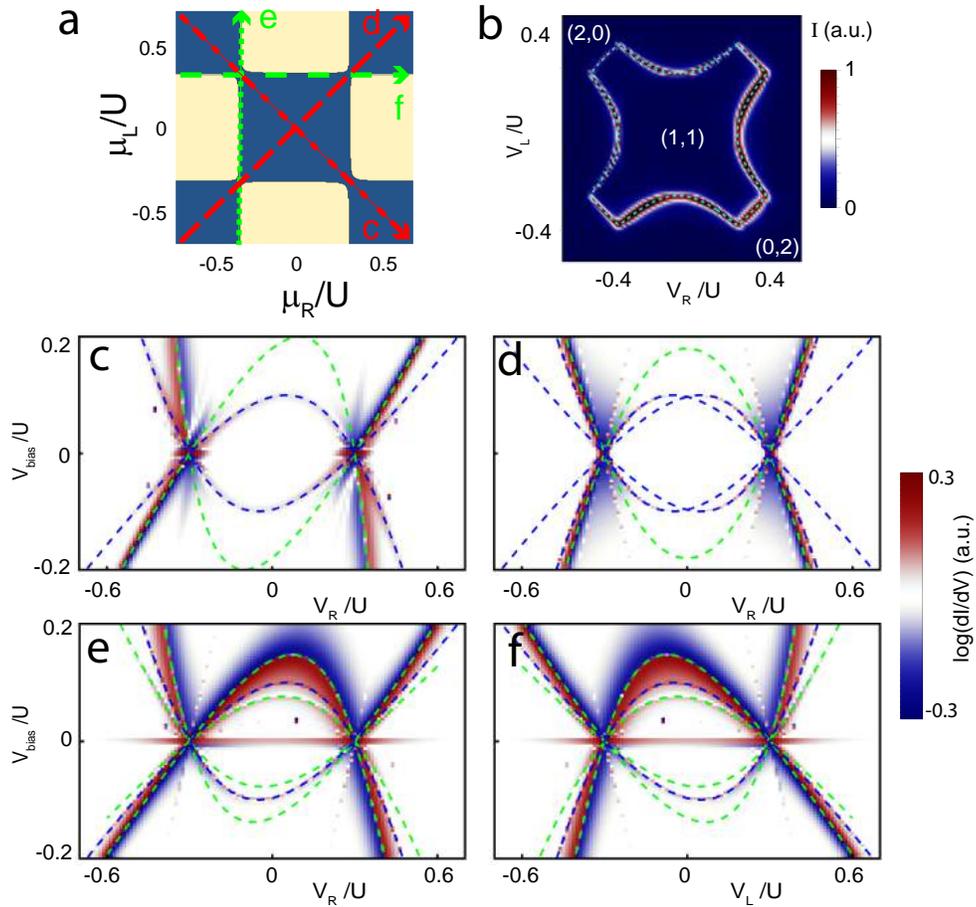

FIG. S12. The diagram (**a**) used to illustrate the line cuts for **b-e** is the same as Fig.S11**a**. **b** is the same as Fig.2c. **c-f** are bias vs. gate plots along cuts depicted in **a**. The green dashed lines indicate the bias voltage at which the levels on the quantum dots come into resonance and the purple dashed lines indicate when the bias voltage exceeds the Andreev level spacing [see Sec. III F for details]. Notice that additional resonances near zero bias are shown inside the loop-like resonances (**e,f**), which is consistent with the experimental observations (Fig.1**a-b**, Fig.S4**d,f**, and Fig.S8**e**).

## M.  The simulated spin map

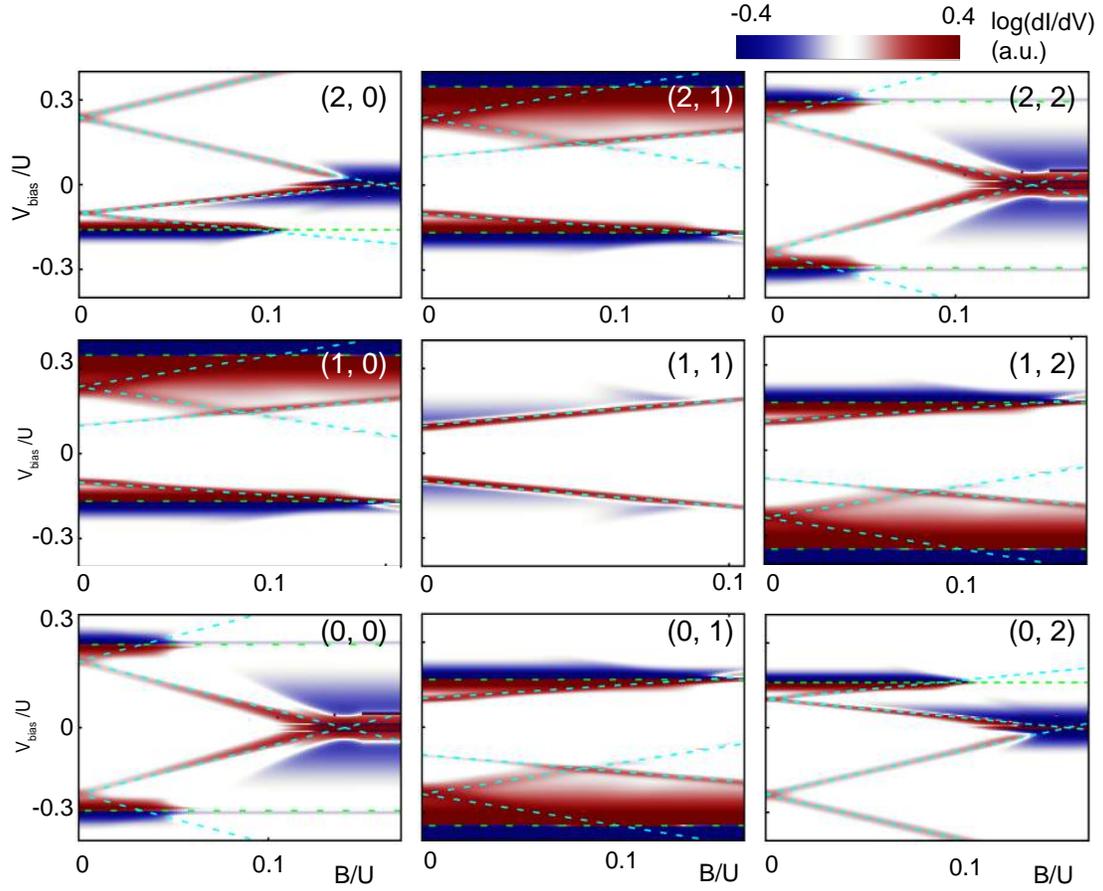

FIG. S13. Simulated magnetic field dependence of the 3x3 configurations around (1, 1). Besides the good fitting between measurements and simulations in each individual configuration, the experimental (Fig.S5) and the simulated maps exhibit impressively similar patterns. First, (0,0) and (2,2) show resonances symmetrically at both biases, while only the negative (positive) branch has high conductance in (2,0) ((0,2)). Second, in (1,0) and (2,1) the kinks occur at the positive bias while in (0,1) and (1,2) the kinks occur at the negative bias. We notice that even with stronger interdot coupling (Fig.S9), such consistency is still valid.

## II. ANDREEV MOLECULES AT ZERO BIAS

In this section we explicitly write down the model Hamiltonians for Andreev atoms and molecules that were used to produce Fig. 1c,d of the main text.

The minimal model for Andreev bound states (Andreev atomic states) in a superconductor-quantum-dot (S-QD) system at zero bias consists of a single level quantum dot coupled to the superconducting reservoir. The model is described by the Hamiltonian

$$H_{\text{S-QD}} = \sum_{\sigma=\{\uparrow,\downarrow\}} \varepsilon_\sigma n_\sigma + U n_\uparrow n_\downarrow - \left( \tau d_\uparrow^\dagger d_\downarrow^\dagger + \text{H.c.} \right), \tag{1}$$

where $\varepsilon_\sigma$ is the energy of an electron with spin $\sigma$ to be on the quantum dot which is set by the gate voltage and the magnetic field, $n_\sigma = d_\sigma^\dagger d_\sigma$ is the number operator for electrons with spin $\sigma$ on the dot, $U$ is the charging energy for double occupancy of the dot, $\tau$ describes the strength of Andreev reflection of electrons from the superconductor and $d_\sigma^\dagger$ and $d_\sigma$ are the electron creation and annihilation operators. The spectrum of this Hamiltonian consists of two spin-singlets ($\sin(\theta)|0\rangle + \cos(\theta)|\uparrow\downarrow\rangle$ and $\cos(\theta)|0\rangle - \sin(\theta)|\uparrow\downarrow\rangle$) and a spin-doublet ($|\uparrow\rangle$ and $|\downarrow\rangle$). The parity of the ground state, as well as the angle $\theta$, is determined by the coupling constants and can be tuned from even to odd to even by varying $\varepsilon_\uparrow$ and $\varepsilon_\downarrow$. The lower of the spin singlet states and the spin doublet states are plotted in Fig. 1c.

To describe Andreev molecular states at zero bias we need to couple two Andreev atoms. The corresponding Hamiltonian becomes

$$H_{\text{S-QD-QD-S}} = \sum_{i=\{L,R\}} \left[ \sum_{\sigma=\{\uparrow,\downarrow\}} \varepsilon_{i,\sigma} n_{i,\sigma} + U_i n_{i,\uparrow} n_{i,\downarrow} - \left( \tau_i d_{i,\uparrow}^\dagger d_{i,\downarrow}^\dagger + \text{H.c.} \right) \right] \tag{2}$$

$$- t \sum_{\sigma=\{\uparrow,\downarrow\}} \left( d_{L,\sigma}^\dagger d_{R,\sigma} + \text{H.c.} \right), \tag{3}$$

where the subscripts $L$ and $R$ stand for the left and the right quantum dots and $t$ is the inter-dot tunneling matrix element. The eigenstates of $H_{\text{S-QD-QD-S}}$ are the Andreev molecular states plotted in Fig. 1d.

In the following section, we discuss a more detailed model of Andreev molecular states at nonzero bias voltages, which describes both Andreev reflections and inter-dot coupling while keeping track of the charging energy of the two superconducting leads.

## III.  DETAILED DESCRIPTION OF OUR TRANSPORT MODEL

### A.  Model Hamiltonian

We use two main ingredients to setup our detailed model of the transport properties of the superconductor-double-dot-superconductor (S-QD-QD-S) device. The first ingredient is a ladder of Andreev molecular states of the S-QD-QD-S system. These states are a result of the interplay of Andreev reflections and hybridization between the quantum dots. The second ingredient is a low, but finite, density of sub-gap quasi-particle states in the leads. The resonant tunneling of these sub-gap quasi-particles into the double dot subsystem is responsible, within our model, for driving transitions between the eigenstates of the S-QD-QD-S system and hence for the experimentally observed sub-gap transport features. To account

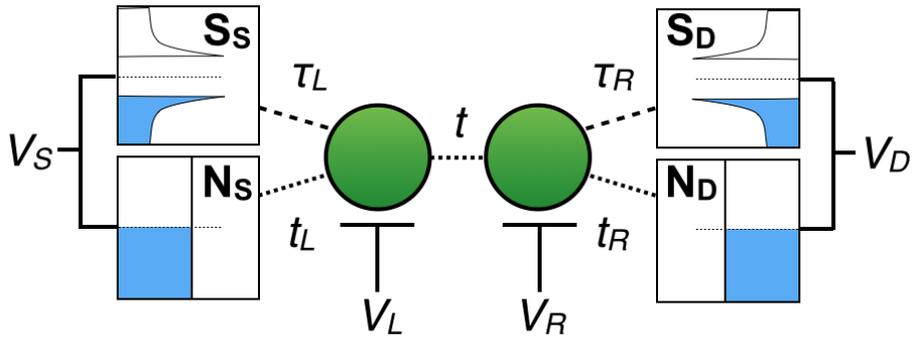

FIG. S14. Hybrid superconductor-double-dot-superconductor system, consisting of an array of two quantum dots tunnel-coupled to superconducting leads. Each lead is modeled as having a standard BCS superconducting (**S**) component and a normal metal (**N**) component. The coupling to the BCS components give rise to Andreev reflection processes, whereas the coupling to the normal components provide low-energy electronic excitations which are responsible for sub-gap transport. The strength of the inter-dot tunneling is set by $t$, while $\tau_L$ and $\tau_R$ ($t_L$ and $t_R$) control the coupling of the left and right dots to the superconducting (normal) components of the left and right leads. The leads are biased by the source (drain) voltages $V_{S(D)}$ and the chemical potential on the left (right) dot is controlled by the side-gate voltage $V_{L(R)}$.

for the presence of sub-gap quasi-particles, we model each lead as having a superconducting component and a normal metal component [see Fig. S14]. The superconducting component is a conventional BCS superconductor with a hard gap $\Delta$, which provides a condensate of

Cooper pairs and drives Andreev reflection processes. The normal component, which we model as a non-interacting Fermi gas with a low density of states at the Fermi surface, provides the low energy electronic excitations that are necessary for sub-gap transport. Finally, to model the application of bias, $V_{\text{bias}}$, we tie the electro-chemical potentials of the two components together and fix them to the applied bias voltage [see Fig. S14].

One of the key features observed in the experimental data are discrete and narrow Andreev bound state-like features. As the strong resonant tunneling of electrons from the normal component of the leads to the quantum dot sub-system tends to broaden the discrete levels of the quantum dot sub-system, we restrict our modeling to the regime where single-electron tunneling (between the quantum dot system and the normal metal component of the leads) is the weakest coupling in the system.

Our transport model is encoded by the Hamiltonian

$$H = H_{\text{QD}} + H_{\text{S}} + H_{T,S} + H_N + H_{T,N}, \tag{4}$$

where $H_{\text{QD}}$ describes the double-dot subsystem, $H_{\text{S}}$ describes the electro-chemical potential energy of the Cooper pairs in the superconducting leads, $H_{T,S}$ describes Andreev reflection, $H_N$ is the Hamiltonian of the normal component of the leads, and $H_{T,N}$ describes the tunneling between the QDs and the normal components of the leads. $H_{\text{QD}}$ is given by

$$H_{\text{QD}} = \sum_{j\sigma} \varepsilon_{j\sigma} n_{j\sigma} + U \sum_j n_{j\uparrow} n_{j\downarrow} - t \sum_\sigma \left( d^\dagger_{R,\sigma} d_{L\sigma} + d^\dagger_{L\sigma} d_{R,\sigma} \right), \tag{5}$$

where $n_{j\sigma} = d^\dagger_{j\sigma} d_{j\sigma}$ is the number operator of the electrons on QD $j = \{L, R\}$ with spin $\sigma$, energy $\varepsilon_{j\sigma}$ (controlled by the electro-chemical potential in quantum dot $j$). The strength of the Coulomb repulsion and of the inter-dot coupling is set by $U$ and $t$, respectively. The model Hamiltonian for the leads is a combination of the superconducting component

$$H_{\text{S}} = \sum_{j\in\{S,D\}} eV_j N_j, \tag{6}$$

and the normal component

$$H_N = \sum_{j\in\{S,D\}} \sum_{\mathbf{k}\sigma} (\xi_{\mathbf{k}} + eV_j) \, c^\dagger_{j\mathbf{k}\sigma} c_{j\mathbf{k}\sigma}, \tag{7}$$

where $j = \{S, D\}$ indicates the source and drain leads, $N_j$ represents the electron number operator for the superconducting component, $c^\dagger_{j\mathbf{k}\sigma}$ ($c_{j\mathbf{k}\sigma}$) creates (annihilates) an electron

with momentum $\mathbf{k}$ and spin $\sigma$ with energy $\xi_{\mathbf{k}}$ in the normal component of lead $j$, and both components are biased by the same voltages $V_j$. The Andreev reflection (i.e. pair tunneling) is described by the Hamiltonian

$$H_{T,S} = -\tau_L S_S^+ d_{L\downarrow} d_{L\uparrow} - \tau_R S_D^+ d_{R\downarrow} d_{R\uparrow} + \text{H.c.}, \tag{8}$$

where the operator $S_j^+$ increases the number of electrons in the superconducting condensate of the $j$-th superconducting lead by two: $N_j \to N_j + 2$. Keeping track of the number of electrons in the superconducting condensates in the two leads is an essential feature of the model that allows us to describe Andreev reflection between the QDs and both leads when there is a voltage difference between the leads[?] . The coupling between the QDs and the normal leads is given by the conventional tunneling Hamiltonian

$$H_{T,N} = -t_L \sum_{\mathbf{k}\sigma} c_{S\mathbf{k}\sigma}^\dagger d_{L\sigma} - t_R \sum_{\mathbf{k}\sigma} c_{D\mathbf{k}\sigma}^\dagger d_{R\sigma} + \text{H.c.}. \tag{9}$$

Here $\tau_j$ and $t_j$ (taken to be real) set the strength of the pair (Andreev reflection) and single-electron tunneling between QD and lead $j$.

As $t_L$ and $t_R$ are the weakest couplings in the system, we call $H_{\text{AMH}} = H_{\text{QD}} + H_{\text{S}} + H_{T,S}$ the Andreev molecular Hamiltonian and treat $H_{T,N}$ as a perturbation to $H_{\text{AMH}}$. That is, the Andreev molecular Hamiltonian gives rise to the Andreev molecular states, and $H_{T,N}$ drives transitions between these states.

### B. Eigenstates of the Andreev molecular Hamiltonian at finite bias

The Andreev molecular Hamiltonian preserves the total electron number $N_T$, total parity, total spin $S_T$, and spin projection $S_z$. Therefore, the Andreev molecular states of the S-QD-QD-S system can be split into subspaces of even and odd parity; the even subspace consists of singlet ($S$) and triplet ($T_{0,\pm}$) Andreev molecular states, whereas the odd parity subspace consists of doublet ($D_\pm$) Andreev molecular states.

In terms of the number of Cooper pairs in the source and drain leads, $N_L$ and $N_R$, what do eigenstates of the Andreev molecular Hamiltonian look like at finite bias? A good analogy are the spatially localized eigenstates of a quantum particle in a tilted washboard potential. Although the ground state corresponds to the particle at the "bottom" of the washboard, there is a whole ladder of eigenstates $\psi_i$, one eigenstate for each lattice site, that

lead to Bloch oscillations. Given the eigenstate $\psi_i$ we can find the state $\psi_{i+1}$ by shifting the wavefunction one lattice site down. Hence the eigenstates can be thought of as forming a ladder, with the rungs labeled by the expectation value for position $\langle x \rangle_i$. Similarly, Andreev molecular states "live" on ladders, with rungs corresponding to the "shift" $s \approx N_L - N_R$ [which will be precisely defined in the next paragraph]. Given an eigenstate on a particular rung, we can obtain the eigenstate on the next rung by shifting a Cooper pair from the left lead to the right lead. For the case of the double quantum dot system there are 16 ladders (4 spin up doublets, 4 spin down doublets, 3 triplets and 5 singlets), which we label by the spin state and "color".

Consider, for example, the singlet Andreev molecular subspace with $N_T = 2N$ electrons. Due to Pauli blockade, there are only five possible ways of electrons occupying the double-dot orbitals, namely $|0,0\rangle, |0,\uparrow\downarrow\rangle, |\uparrow\downarrow,0\rangle, |\uparrow\downarrow,\uparrow\downarrow\rangle$ and $(|\uparrow,\downarrow\rangle - |\downarrow,\uparrow\rangle)/\sqrt{2}$. When coupled to the superconducting leads via Andreev reflection, those five double-dot, singlet states hybridize with the bare states of the superconductors $|N_S, N_D\rangle$, which represent a given distribution of Cooper pairs between the leads. Thus, we can generate any state in this subspace from five reference states of the S-QD-QD-S system, such as $|N,0,0,N\rangle, |N,0,\uparrow\downarrow,N-2\rangle, |N-2,\uparrow\downarrow,0,N\rangle, |N-2,\uparrow\downarrow,\uparrow\downarrow,N-2\rangle$ and $(|N,\uparrow,\downarrow,N-2\rangle - |N,\downarrow,\uparrow,N-2\rangle)/\sqrt{2}$, by transferring Cooper pairs from one lead to the other using the transfer operators $T^{\pm}|N_S, \mathrm{QD_L}, \mathrm{QD_R}, N_D\rangle = |N_1 \pm 2, \mathrm{QD_L}, \mathrm{QD_R}, N_2 \mp 2\rangle$. This is possible since all remaining states of the span correspond to one of the reference states, but with a different Cooper pair configuration. By linearity, the same considerations apply to the eigenstates. Hence, the whole ladder of singlet Andreev molecular states $|S^{(c,s)}\rangle \equiv |N_T = 2N, S_T = 0, S_z = 0, c, s\rangle$ can be constructed from the five reference eigenstates, which we refer to by the "color" quantum number ($c = 1, 2, \ldots, 5$). The number of unique color eigenstates corresponds to the number of unique Andreev molecular states. As a result, the triplet and doublet subspaces can be generated from sets of three and eight color eigenstates, respectively. Here we also introduce the "shift" quantum number ($s = 0, \pm 1, \pm 2, \ldots$), defined as the number of times one needs to apply $T^{\pm}$ to a reference eigenstate to generate an eigenstate with a different Cooper pair configuration.

We define the $s = 0$ reference shift as the eigenstates whose maximum components show minimum Cooper pair imbalance between the leads. We remark that this definition is arbitrary, and alternative definitions should not effect physical results. Note that the

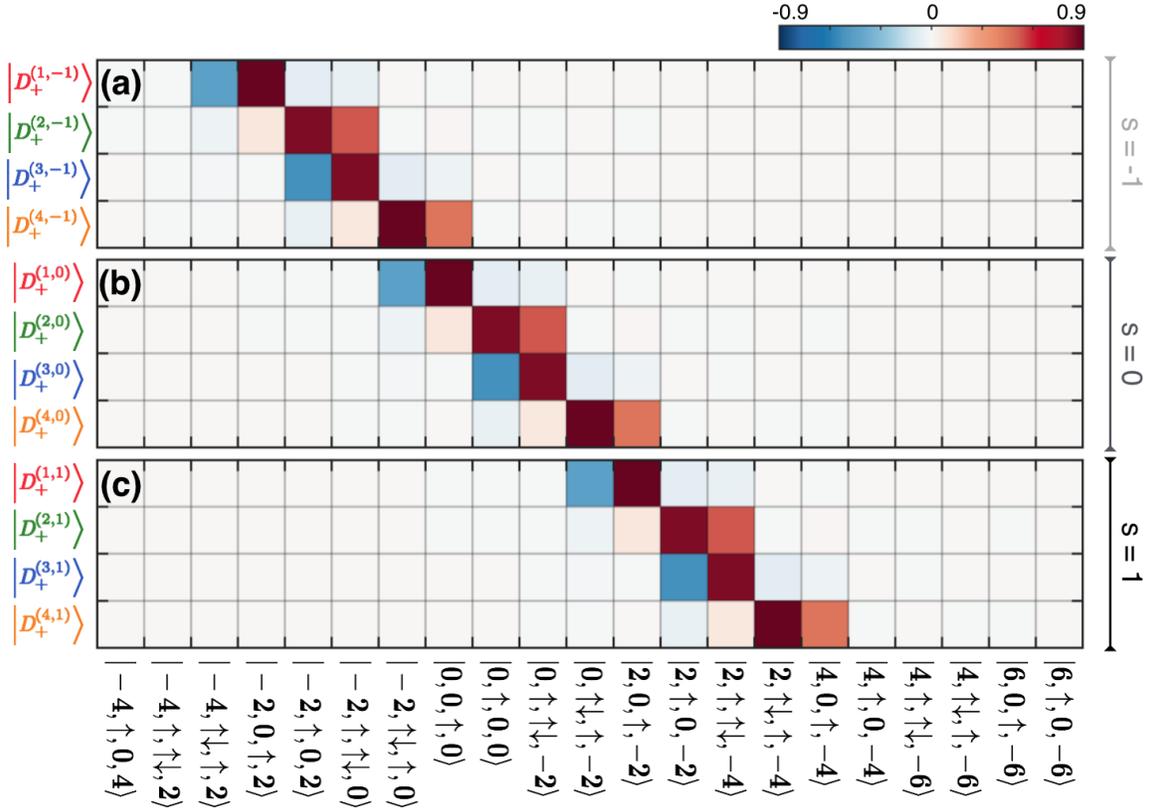

FIG. S15. Ladder of Andreev molecular states for the doublet subspace $\left| D_+^{(c,s)} \right\rangle \equiv |N_T = 2N + 1, S = 1/2, S_z = 1/2, c, s\rangle$ for $V_{\text{bias}} = V_D - V_S = 0.2\Delta/e$, $\tau_{L(R)} = 0.8\Delta$, $t = 0.01\Delta$. We show the set of four color eigenstates corresponding to three shifts (a) $s = -1$, (b) $s = 0$, and (c) $s = 1$. The $s = 0$ reference shift corresponds to the set of color states whose maximum components have minimum Cooper pair imbalance between the leads. From those states, we generate the set of states for the subsequent $s = 1$ ($s = -1$) shift by transferring one Cooper pair from the drain (source) lead to the source (drain) lead. For convenience, we choose to count electrons relative to $N$, which is equivalent to set $N = 0$.

eigenenergies for non-zero shifts ($s \neq 0$) can then be easily obtained from the relation

$$E_{N_T,S_z}^{c,s} = E_{N_T,S_z}^{c,0} + 2s\,e(V_S - V_D). \tag{10}$$

As an example, we show in Fig. S15 the $s = 0$ color states for the $D_+$ subspace for different bias voltages. At larger bias voltages, the eigenstates are well localized in Hilbert space, showing a narrow distribution of Cooper pairs. As the bias voltage decreases towards zero, the number of Cooper pairs are allowed to fluctuate and, as a result, the eigenstates spread.

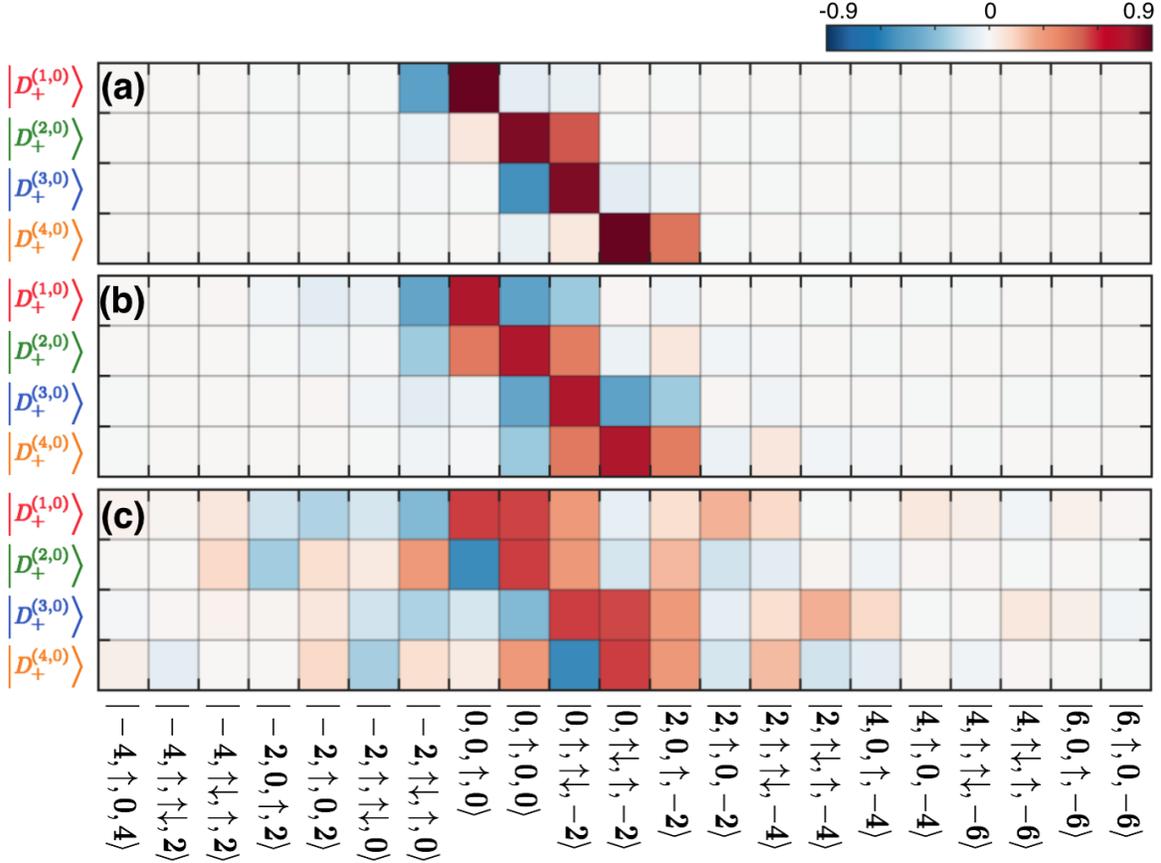

FIG. S16. Reference $s = 0$ color states for the doublet Andreev molecular subspace $\left|D_+^{(c,0)}\right\rangle \equiv |N_T = 2N+1, S = 1/2, S_z = 1/2, c, s = 0\rangle$ for (a)–(c) $eV_{\text{bias}}/\Delta = 0.2, 0.02, 0.002$, showing the spreading of the probability amplitudes at low bias voltages.

## C. Classical master equation

To describe the experimentally observed sub-gap transport through the S-QD-QD-S device, we now consider the effects of the coupling to the normal component of the leads. We describe the state of the S-QD-QD-S device by the probability distribution $P$, which gives the probability of finding the system in a particular eigenstate $|N_T, S_T, S_z, c, s\rangle$ of the Andreev molecular Hamiltonian. The S-QD-QD-S system is pushed out of equilibrium by a nonzero source-drain bias voltage. Energy is dissipated by single electrons tunneling from the quantum dots to the normal components of the leads. Such incoherent processes drive transitions between Andreev molecular subspaces of different parity, as illustrated in Fig. S17.

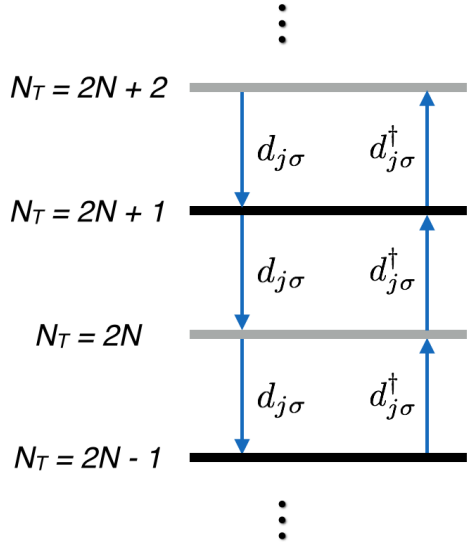

FIG. S17. Transitions between the even and odd Andreev molecular subspaces of the S-QD-QD-S system driven by single electron tunneling between the normal leads and the double dot subsystem. Depending on the spin of the exchanged electron, these transitions couple doublet states to either singlet or triplet states.

We write a classical master equation that accounts for the transitions between the various Andreev molecular levels. Depending on the spin of the exchanged electron, these transitions couple doublet states to either singlets or triplets. As we are interested in describing the transport dynamics in the long time limit, the non-equilibrium probability distribution $P$ is given by the steady state solution of the rate equation

$$\frac{\mathrm{d}P(\mathbf{n})}{\mathrm{dt}} = \sum_{\mathbf{m}} \left( \Gamma_{\mathbf{n}\leftarrow\mathbf{m}} P(\mathbf{m}) - \Gamma_{\mathbf{m}\leftarrow\mathbf{n}} P(\mathbf{n}) \right), \tag{11}$$

where the first (second) term represents the probability of tunneling into (out of) state $|\mathbf{n}\rangle \equiv |N_T, S_T, S_z, c, s\rangle$ and $\Gamma_{\mathbf{n}\leftarrow\mathbf{m}}$ are the transition rates between levels $\mathbf{m}$ and $\mathbf{n}$ due to the exchange of one electron with the normal leads[3]. Specifically, if the transition rate $\Gamma_{\mathbf{n}\leftarrow\mathbf{m}}$ results from the addition of an electron to the S-QD-QD-S system, it is given by

$$\Gamma_{\mathbf{n}\leftarrow\mathbf{m}}^{(\text{gain})} = 2\pi \sum_{j,\sigma} t_j^2 |\langle \mathbf{n}| d_{j,\sigma}^\dagger |\mathbf{m}\rangle|^2 n_F(E_{\mathbf{n}} - E_{\mathbf{m}} - eV_j), \tag{12}$$

whereas if it results in the loss of an electron to the normal leads, we have

$$\Gamma_{\mathbf{n}\leftarrow\mathbf{m}}^{(\text{loss})} = 2\pi \sum_{j,\sigma} t_j^2 |\langle \mathbf{n}| d_{j,\sigma} |\mathbf{m}\rangle|^2 \left(1 - n_F(E_{\mathbf{m}} - E_{\mathbf{n}} - eV_j)\right). \tag{13}$$

Here $n_F$ represents the Fermi-Dirac distribution, which gives the probability to find an electron in the normal leads, and $\{E_\mathbf{m}\}$ represent the eigenenergies of the S-QD-QD-S system. The rate equation (11) takes into account all possible single-electron transitions between Andreev molecular states. However, as we show below, we can use the symmetries of the Andreev molecular Hamiltonian to effectively reduce Eq. (11) to involve only transitions between two subspaces of opposite parity, containing a total of $2N$ and $2N+1$ electrons ($N \gg 1$).

As shown in Fig. S17, the $2N$ and $2N+1$ subspaces are directly coupled by transitions involving either the addition of an electron to the $2N$ subspace or the removal of an electron from the $2N+1$ subspace. Using Eqs. (12) and (13), those rates are given by

$$\Gamma^{(\text{gain})}_{\alpha_O \leftarrow \beta_E} = 2\pi \sum_{j,\sigma} t_j^2 |\langle 2N+1, \alpha_O, c', s'| d_{j\sigma}^\dagger |2N, \beta_E, c, s \rangle|^2 n_F\big(E^{c',s'}_{2N+1,\alpha_O} - E^{c,s}_{2N,\beta_E} - eV_j\big),$$
(14)

$$\Gamma^{(\text{loss})}_{\alpha_E \leftarrow \beta_O} = 2\pi \sum_{j,\sigma} t_j^2 |\langle 2N, \alpha_E, c', s'| d_{j\sigma} |2N+1, \beta_O, c, s \rangle|^2 \Big(1 - n_F\big(E^{c,s}_{2N+1,\beta_O} - E^{c',s'}_{2N,\alpha_E} - eV_j\big)\Big).$$
(15)

For simplicity of notation, here and henceforth we label the spin subspaces by $\alpha_{E(O)}, \beta_{E(O)} = S, T_{0,\pm}(D_\pm)$, where the $E, O$ subscripts emphasize that these are transitions between the $N_T = 2N$ (even) and $N_T = 2N+1$ (odd) Andreev molecular subspaces of the S-QD-QD-S system.

We refer to transitions described by Eqs. (14) and (15) as type 1 transitions. Type 2 transitions connect the $2N$ and $2N+1$ subspaces to the $2N-1$ and $2N+2$ subspaces (see Fig. S17). Type 2 transitions can be mapped back onto the $2N$ and $2N+1$ subspaces because when the number of Cooper pairs is changed by one on a lead at fixed bias voltage, within our model, the eigenenergies are trivially shifted according to the change in electrochemical potential energy, i.e., $E^{c,s}_{N_T \pm 2,\alpha} = E^{c,s}_{N_T,\alpha} \pm 2eV_j$. Thus, by using this relation and the operators $S^+$ and $S^-$, we can write the following identities:

$$|\langle 2N+1, \alpha', c', s'| d_{j\sigma} |2N+2, \alpha, c, s \rangle|^2 \Big(1 - n_F\big(E^{c,s}_{2N+2,\alpha} - E^{c',s'}_{2N+1,\alpha'} - eV_j\big)\Big)$$
$$= |\langle 2N+1, \alpha', c', s'| d_{j\sigma} S^+ |2N, \alpha, c, s \rangle|^2 \Big(1 - n_F\big(E^{c,s}_{2N,\alpha} - E^{c',s'}_{2N+1,\alpha'} + eV_j\big)\Big), \quad (16)$$

$$|\langle 2N+2, \alpha, c', s'| d_{j\sigma}^\dagger |2N+1, \beta, c, s \rangle|^2 n_F\big(E^{c',s'}_{2N+2,\alpha} - E^{c,s}_{2N+1,\beta} - eV_j\big)$$
$$= |\langle 2N, \alpha, c', s'| S^- d_{j\sigma}^\dagger |2N+1, \beta, c, s \rangle|^2 n_F\big(E^{c',s'}_{2N,\alpha} - E^{c,s}_{2N+1,\beta} + eV_j\big), \quad (17)$$

Equations (16) and (17) thus show how transitions of type 2 can be effectively mapped onto a transition between the $2N$ and $2N+1$ subspaces. Note that those transitions are driven by the exchange of a single electron with the normal leads together with the creation or annihilation of a Cooper pair on the superconducting component of the same lead. The rates for type-2 transitions are then given by

$$\tilde{\Gamma}^{(\text{gain})}_{\alpha_E \leftarrow \beta_O} = 2\pi \sum_{j,\sigma} t_j^2 |\langle 2N, \alpha_E, c', s'| \, S^- d^\dagger_{j\sigma} \, |2N+1, \beta_O, c, s\rangle \, |^2 n_F\big(E^{c',s'}_{2N,\alpha_E} - E^{c,s}_{2N+1,\beta_O} + eV_j\big),$$
(18)

$$\tilde{\Gamma}^{(\text{loss})}_{\alpha_O \leftarrow \beta_E} = 2\pi \sum_{j,\sigma} t_j^2 |\langle 2N+1, \alpha_O, c', s'| \, S^+ d_{j\sigma} \, |2N, \beta_E, c, s\rangle \, |^2 \big(1 - n_F\big(E^{c,s}_{2N,\beta_E} - E^{c',s'}_{2N+1,\alpha_O} + eV_j\big)\big),$$
(19)

which describe either the removal of an electron from an even parity state or the addition of an electron to an odd eigenstate.

It is easy to generalize this mapping to all other $N_T$ subspaces and show that any single-electron transition rate are of type 1 or 2 and, hence, can be calculated from Eqs. (14), (15), (18), or (19). Those effective transitions between the $2N$ and $2N+1$ subspaces are illustrated in Fig. S18. In this way, we reduce Eq. (11) to a single even ($2N$) and odd ($2N+1$) subspaces. From now on, we simply refer to those subspaces as even and odd.

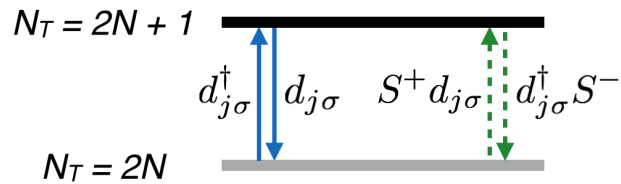

FIG. S18. Effective transitions between the even ($N_T = 2N$) and odd ($N_T = 2N+1$) Andreev molecular subspaces of the S-QD-QD-S system. The solid, blue arrows represent transitions of type 1, which involve the exchange of an electron between the normal leads and the quantum dots (see Eqs. (14) and (15)). Transitions of type 2 (dashed, green arrows), on the other hand, are driven by the exchange of an electron between the dots and the normal lead, but followed by the creation or annihilation of a Cooper pair in the superconducting component of the same lead (see Eqs. (18) and (19)).

We obtain the steady-state solution of Eq. (11) from the eigenvalue equation

$$
\begin{pmatrix}
M_S^{\text{out}} & M_{S \leftarrow D_+}^{\text{in}} & M_{S \leftarrow D_-}^{\text{in}} & 0 & 0 & 0 \\
M_{D_+ \leftarrow S}^{\text{in}} & M_{D_+}^{\text{out}} & 0 & M_{D_+ \leftarrow T_+}^{\text{in}} & M_{D_+ \leftarrow T_0}^{\text{in}} & M_{D_+ \leftarrow T_-}^{\text{in}} \\
M_{D_- \leftarrow S}^{\text{in}} & 0 & M_{D_-}^{\text{out}} & M_{D_- \leftarrow T_+}^{\text{in}} & M_{D_- \leftarrow T_0}^{\text{in}} & M_{D_- \leftarrow T_-}^{\text{in}} \\
0 & M_{T_+ \leftarrow D_+}^{\text{in}} & M_{T_+ \leftarrow D_-}^{\text{in}} & M_{T_+}^{\text{out}} & 0 & 0 \\
0 & M_{T_0 \leftarrow D_+}^{\text{in}} & M_{T_0 \leftarrow D_-}^{\text{in}} & 0 & M_{T_0}^{\text{out}} & 0 \\
0 & M_{T_- \leftarrow D_+}^{\text{in}} & M_{T_- \leftarrow D_-}^{\text{in}} & 0 & 0 & M_{T_-}^{\text{out}}
\end{pmatrix}
\begin{pmatrix}
\vec{P}_S \\
\vec{P}_{D_+} \\
\vec{P}_{D_-} \\
\vec{P}_{T_+} \\
\vec{P}_{T_0} \\
\vec{P}_{T_-}
\end{pmatrix} = 0, \quad (20)
$$

where the matrices $M_{\alpha \leftarrow \beta}^{\text{in}} = \Gamma_{\alpha \leftarrow \beta} + \tilde{\Gamma}_{\alpha \leftarrow \beta}$ and $M_\alpha^{\text{out}} = -\sum_\beta \left( \Gamma_{\beta \leftarrow \alpha} + \tilde{\Gamma}_{\beta \leftarrow \alpha} \right)$ describe the influx and outflux of probability of subspace $\alpha$ (with $\alpha, \beta = S, D_\pm, T_{0,\pm}$). Note that the vectors $\vec{P}_\alpha$ have dimension $d_c^\alpha d_s^\alpha$, where $d_c^\alpha$ ($d_s^\alpha$) is the number of color (shift) states in subspace $\alpha$. Similarly, $M_\alpha^{\text{out}}$ and $M_{\alpha \leftarrow \beta}^{\text{in}}$ are matrices of dimension equal to $d_c^\alpha d_s^\alpha \times d_c^\alpha d_s^\alpha$ and $d_c^\beta d_s^\alpha \times d_c^\beta d_s^\beta$.

## D. Steady-state Current

The steady-state current is obtained from the rate at which electrons go through the S-QD-QD-S device and is given by

$$
I = -e \sum_{\alpha, \beta} \left( \gamma_{R, \alpha \leftarrow \beta}^{(\text{gain})} - \gamma_{R, \alpha \leftarrow \beta}^{(\text{loss})} \right) \vec{P}_\beta \quad (21)
$$

where the matrices $\gamma_{R, \alpha \leftarrow \beta}^{(\text{gain})}$ and $\gamma_{R, \alpha \leftarrow \beta}^{(\text{loss})}$ (of dimension $d_c^\alpha d_s^\alpha \times d_c^\beta d_s^\beta$) provide the current rates for transitions to the odd subspace, which are given by

$$
\gamma_{R, \alpha_O \leftarrow \beta_E}^{(\text{gain})} = 2\pi t_j^2 \left| \left\langle \alpha_O^{(c',s')} \middle| d_{j\sigma}^\dagger \middle| \beta_E^{(c,s)} \right\rangle \right|^2 n_F\!\left( E_{\alpha_O}^{c',s'} - E_{\beta_E}^{c,s} - eV_j \right) (2s' - 2s - 1), \quad (22)
$$

$$
\gamma_{R, \alpha_O \leftarrow \beta_E}^{(\text{loss})} = 2\pi t_j^2 \left| \left\langle \alpha_O^{(c',s')} \middle| S^+ d_{j\sigma} \middle| \beta_E^{(c,s)} \right\rangle \right|^2 \left( 1 - n_F\!\left( E_{\beta_E}^{c,s} - E_{\alpha_O}^{c',s'} + eV_j \right) \right) (2s' - 2s - 1), \quad (23)
$$

and for transitions to the even subspace, which are given by

$$
\gamma_{R, \alpha_E \leftarrow \beta_O}^{(\text{gain})} = 2\pi t_j^2 \left| \left\langle \alpha_E^{(c',s')} \middle| S^- d_{j\sigma}^\dagger \middle| \beta_O^{(c,s)} \right\rangle \right|^2 n_F\!\left( E_{\alpha_E}^{c',s'} - E_{\beta_O}^{c,s} + eV_j \right) (2s' - 2s - 1), \quad (24)
$$

$$
\gamma_{R, \alpha_E \leftarrow \beta_O}^{(\text{loss})} = 2\pi t_j^2 \left| \left\langle \alpha_E^{(c',s')} \middle| d_{j\sigma} \middle| \beta_O^{(c,s)} \right\rangle \right|^2 \left( 1 - n_F\!\left( E_{\beta_O}^{c,s} - E_{\alpha_E}^{c',s'} - eV_j \right) \right) (2s' - 2s - 1). \quad (25)
$$

### E.   Numerical simulations

We solve the master equation (20) and compute the current (21) numerically, by first finding the set of even and odd eigenstates of the S-QD-QD-S Hamiltonian (4) via exact diagonalization. As discussed above, we restrict this calculation to even and odd subspaces with a total of $2N$ and $2N+1$ electrons. Note that because of the conservation of the total electron number, the exact value of $N$ only sets an overall offset and, hence, $N$ can be taken as an arbitrary parameter. After diagonalizing $H$, we select the reference color eigenstates and eigenenergies for each subspace. Together with the Cooper pair transfer operators $T^{\pm}$, we then construct the ladders of Andreev molecular states, whose energies are calculated from Eq. (10). We use the ladders of Andreev states to compute the transition probabilities between the even and odd subspaces and their respective Fermi electron (hole) occupation probabilities on the normal leads. Finally, this allow us to calculate the transition rates (14), (15), (18), (19), and (22)–(25) and then to construct and solve both the master equation (20) and the current equation (21).

### F.   Origin of features in conduction

Let us now consider the requirements for transport in the S-QD-QD-S system with weak inter-dot coupling [see Fig. S14], i.e. in the regime where the two quantum dots are cannot be considered to be merged into a single dot. We require two conditions: (1) the source-drain bias $V_{\text{bias}}$ must be sufficiently high to ensure that electrons can resonantly tunnel from the normal component of the lead into (or out of) the quantum dots, and (2) there must be a resonant process for moving electrons between the dots. Suppose that the ground and excited state energies of the two S-QD subsystems are $E_{\{g,x\},\{L,R\}}$. Then, condition (1) requires that $V_S \geq E_{x,L} - E_{g,L}$ or $V_D \geq E_{x,R} - E_{g,R}$ while condition (2) requires that $E_{x,L} + E_{g,R} = E_{g,L} + E_{x,R}$. In both cases, multiple Andreev reflections will result in moving electrons from the source lead (S) to the drain lead (D), and must be balanced by adding $2nV_{\text{bias}}$ to the energy conservation equations, where $n$ is the number of pairs being moved.

Enforcing these two simple requirements reproduces almost all the qualitative features observed experimentally [see the dashed lines in Fig. 4c,f,i of the main text and Figs. S12,S13 of the supplement]. The one exception, is the asymmetry of the charge stability diagram.

In order to capture this asymmetry it is necessary to include the splitting of the singlet and triplet states, which requires a more detailed model that describes both Andreev reflections and inter-dot coupling while keeping track of the charging energy of the two superconducting leads.



\end{document}